# Quantum percolation theory for dynamic propagation connectivity of transport networks

Junxiang Xu[1], Chence Niu[2,*], Divya Jayakumar Nair[1], Vinayak Dixit[1]

*1. Research Centre for Integrated Transport Innovation (rCITI), School of Civil and Environmental Engineering, The University of New South Wales, Kensington, UNSW Sydney, NSW, 2052, Australia*

*2. Guangdong Basic Research Center of Excellence for Ecological Security and Green Development, Key Laboratory for City Cluster Environmental Safety and Green Development of the Ministry of Education, School of Ecology, Environment and Resources, Guangdong University of Technology, Guangzhou, 510006, China*

**corresponding author*:** niuchence@163.com

**Abstract:** Connectivity degradation in transport networks under structural disturbance is a central problem in network resilience research. Existing methods rely mainly on percolation theory and topological connectivity measures. They focus on whether paths exist and whether connected components fragment. These approaches cannot capture functional degradation where network topology remains intact but propagation ability has already declined substantially. This paper introduces quantum percolation theory into transport network connectivity analysis and proposes Dynamic Propagation Connectivity (DPC) as a new measure that characterises network propagation ability under disturbance. By mapping a transport network under disturbance into a propagation operator system, this paper establishes a spectral analysis framework for DPC and defines the time-averaged participation index as its core quantification. This paper provides a series of rigorous theoretical results. DPC remains constant under homogeneous disturbance and degrades under heterogeneous disturbance. This paper establishes a quantitative relationship between the degradation rate, the minimum eigenvalue spacing of the propagation operator, and heterogeneous deviation strength. This paper proves a separation theorem between DPC and algebraic connectivity. It derives an analytical expression for DPC and a second-order perturbation approximation on the ring graph. Numerical experiments on three transport benchmark networks verify all theoretical conclusions and confirm degradation monotonicity, separation from algebraic connectivity, and degradation amplification by network size. This paper provides a theoretical framework for transport network resilience assessment that goes beyond topological connectivity.

## 1. Introduction

Transport networks support the normal operation of urban economies and the coordination of regional transport systems (Bell and Iida, 1997). Their reliability and connectivity affect daily travel, emergency evacuation, logistics services, and the continuity of social and economic activities (Reggiani et al., 2015). Unexpected events can damage transport infrastructure, cause nodes and links to fail, reduce overall accessibility, increase travel costs, and leave some areas isolated (Xu and Chopra, 2023). Understanding how transport networks behave under structural disturbance and how their connectivity changes is therefore important for network planning and resilience enhancement. Existing research analyses this problem mainly from the perspective of topological connectivity. Topological connectivity examines whether reachable paths exist between nodes and how connected components evolve under disturbance (Li et al., 2021). These approaches have achieved considerable progress in identifying network fragmentation and phase transition thresholds. However, transport networks in practice often exhibit a phenomenon where the network remains connected in the topological sense while its ability to carry and propagate flows has already declined substantially (Cats and Jenelius, 2018). This indicates that topological connectivity alone does not fully describe the functional state of a transport network under disturbance. This paper re-examines the problem from the perspective of propagation connectivity. Propagation connectivity concerns whether a propagation process can sustain and spread across the network, rather than whether paths exist. When structural disturbance causes connection strengths to decay unevenly, the propagation state may concentrate towards a subset of nodes and its spatial extent may contract, even though the network topology remains intact. This paper establishes a mathematical framework for propagation connectivity based on quantum percolation theory and analyses the conditions, rate, and pattern of propagation degradation from the perspective of spectral structure.

### *1.1 Percolation and topological connectivity in transport networks*

Over the past decade, methods based on percolation theory have become a common framework for analysing the robustness and vulnerability of complex networks. The basic idea is to remove network elements step by step following a given rule and to track how indicators such as the size of the giant connected component and the share of connected nodes change to identify critical thresholds and phase transitions (Li et al., 2021).

Percolation studies in transport networks started from structural failure analysis. Li et al. (2015) used real-time traffic data from the Beijing road network to propose the concept of traffic percolation. They showed how the urban network shifts from global connectivity to fragmentation during its daily evolution and demonstrated that improving a small number of key bottleneck links can significantly enhance overall accessibility. Wang et al. (2019) proposed a road percolation failure model that accounts for the effect of terrain height, addressing the limitation of traditional planar network frameworks in capturing the three-dimensional nature of flooding. Hamedmoghadam et al. (2021) introduced heterogeneous demand flows and capacities into the percolation framework. Their study revealed the locations of bottlenecks in infrastructure networks and demonstrated how heterogeneity can increase or reduce the vulnerability of network connectivity.

Recent studies have further strengthened the link between percolation theory and the spread of traffic congestion. Chen et al. (2024) proposed a method for predicting potential network disruption and cascading failure in advance and showed that congestion spreads more strongly when network pressure rises above the traffic percolation threshold. Yao et al. (2025) developed a percolation-based robustness assessment and an SIR-based congestion spreading method that achieved high accuracy in predicting global traffic resilience. Zhu et al. (2025) reviewed the percolation behaviour of rail, bus, and integrated transport networks under link failures and highlighted that percolation indicators have become important tools for assessing structural vulnerability in transport networks.

In terms of topological connectivity measures, algebraic connectivity, defined as the second smallest eigenvalue of the Laplacian matrix, is a classical indicator of global network connectivity strength. Mattsson and Jenelius (2015) reviewed transport network vulnerability and resilience research and highlighted the central role of connectivity measures in resilience assessment. Zhang et al. (2015) analysed the role of network topology in transport network resilience. Cats and Jenelius (2018) showed that partial capacity degradation can reduce network performance without causing topological disconnection. Gu et al. (2023) proposed a vulnerability envelope framework that analysed how multiple link disruptions affect the upper and lower bounds of network performance.

These methods share a common feature. They treat connectivity as a topological property and focus on whether paths exist, whether connected components fragment, and whether algebraic connectivity remains positive. When the network topology stays intact but connection strengths decay unevenly across links, topological connectivity measures cannot capture functional degradation at the propagation level. This limitation motivates the search for an analytical framework that can characterise propagation processes.

*1.2 Quantum percolation and localisation in disordered networks*

Quantum percolation originated in solid-state physics and the study of disordered systems. Anderson (1958) established the theory of localisation in disordered systems and proved that wave functions become spatially confined and diffusion vanishes under sufficient structural disorder. This phenomenon, known as Anderson localisation, forms the foundation of quantum percolation theory. Shapir et al. (1982) further investigated the relationship between localisation and quantum percolation. Avishai and Luck (1992) analysed the ballistic conductance properties of quantum percolation on a lattice of wires. Trugman (1983) connected localisation, percolation, and the quantum Hall effect and established a theoretical bridge among the three.

In the area of quantum propagation on networks, Mülken and Blumen (2011) provided a systematic theoretical review of continuous-time quantum walks on complex networks, established mathematical models for coherent transport, and identified localisation as a central feature of propagation in disordered networks. Schubert et al. (2005) studied localisation effects in quantum percolation and characterised the transition of propagation states from extended to localised behaviour. Biamonte et al. (2019) reviewed the theoretical development from classical networks to quantum networks and discussed the decisive role of spectral structure in quantum network propagation. Xu et al. (2021) experimentally verified quantum transport behaviour on fractal networks and demonstrated the influence of network topology on propagation properties. De Domenico et al. (2016) analysed the physical mechanisms of spreading processes on multilayer networks. Maciel et al. (2020) studied quantum transport on generalised scale-free networks and revealed the influence of network structural parameters on propagation extensibility. Dikopoltsev et al. (2022) experimentally observed Anderson localisation beyond the spectrum of the disorder.

Within the quantum percolation framework, connectivity depends on wave function propagation ability. When structural disorder reaches a certain level, propagation states become localised and propagation ability drops sharply, even though the network remains connected in the topological sense (Schubert et al., 2005). The propagation operator spectral structure, consisting of eigenvalues and eigenvectors, determines whether propagation states are extended or localised. This theory has accumulated extensive numerical and experimental evidence in physics. However, rigorous mathematical results for network connectivity analysis remain limited. Specifically, the conditions under which propagation connectivity degrades, the network structural parameters that control the degradation rate, and the theoretical relationship between propagation connectivity and classical topological connectivity measures remain open questions.

*1.3 Objectives and contributions*

This paper introduces quantum percolation theory into transport network connectivity analysis and establishes a rigorous mathematical framework for propagation connectivity. The focus is on providing theoretical results at the level of theorems and propositions with rigorous mathematical proofs. The main contributions of this paper are as follows.

(1) This paper maps a transport network under disturbance into a propagation operator system through quantum

percolation and defines a spectral analysis framework for propagation connectivity. It proves the rigorous conditions under which propagation connectivity remains constant under homogeneous disturbance and degrades under heterogeneous disturbance.

(2) This paper establishes the theoretical relationship between propagation connectivity and algebraic connectivity and proves that the two can separate. A positive algebraic connectivity does not guarantee that propagation connectivity remains at a high level.

(3) This paper derives an analytical expression for propagation connectivity on the ring graph and obtains a second-order approximation of degradation through perturbation expansion.

(4) This paper verifies all theoretical conclusions through numerical experiments on three transport benchmark networks.

The remainder of this paper is organised as follows. Section 2 formulates the problem and defines the network representation, propagation operator, and propagation connectivity measures. Section 3 presents the theoretical analysis, covering fundamental properties, localisation and degradation, comparison with classical measures, and analytical characterisation on benchmark topologies. Section 4 reports the numerical experiment results. Section 5 discusses the methodological contributions and limitations. Section 6 concludes the paper.

## 2. Problem formulation

*2.1 Transport network under structural disturbance*

Consider a directed transport network $G=(V,E,A)$. The node set $V=\{1,2,...,N\}$ represents traffic zones, intersections, or functional areas. The directed edge set $E\subseteq V\times V$ describes physical or operational connections between nodes. The association matrix $A=\left[a_{ij}\right]$ captures the effective connection strength from node $i$ to node $j$, where $a_{ij}\geq 0$. A positive value $a_{ij}>0$ reflects the functional carrying capacity of the corresponding connection. When $a_{ij}=0$, the connection makes no effective contribution to propagation across the network.

To characterise the impact of structural disturbance on the network, this study introduces a disturbance intensity parameter $\theta\in[0,1]$. The value $\theta=0$ represents the undisturbed state, and $\theta=1$ represents the extreme disturbance state. Under disturbance, the connection strength of each link is:

$$a_{ij}(\theta)=a_{ij}^{0}\cdot\phi_{ij}(\theta) \tag{1}$$

where $a_{ij}^{0}$ denotes the initial connection strength of link $(i,j)$, and $\phi_{ij}(\theta)\in[0,1]$ represents the disturbance decay function. This function satisfies four conditions: $\phi_{ij}(0)=1,\phi_{ij}(1)=1,\phi_{ij}(\theta)$ decreases monotonically with respect to $\theta$, and $\phi_{ij}(\theta)$ is continuous on $[0,1]$.

The impact of structural disturbance on the network falls into two categories. The first category is homogeneous disturbance, where all links share the same decay function $\phi(\theta)$ and experience the same degree of degradation. The second category is heterogeneous disturbance, where each link has an independent decay function $\phi_{ij}(\theta)$ and the degree of degradation varies across links. This distinction forms the basis for the theoretical analysis in subsequent sections.

Figure 1 illustrates the transport network representation under the undisturbed state and two types of structural disturbance.

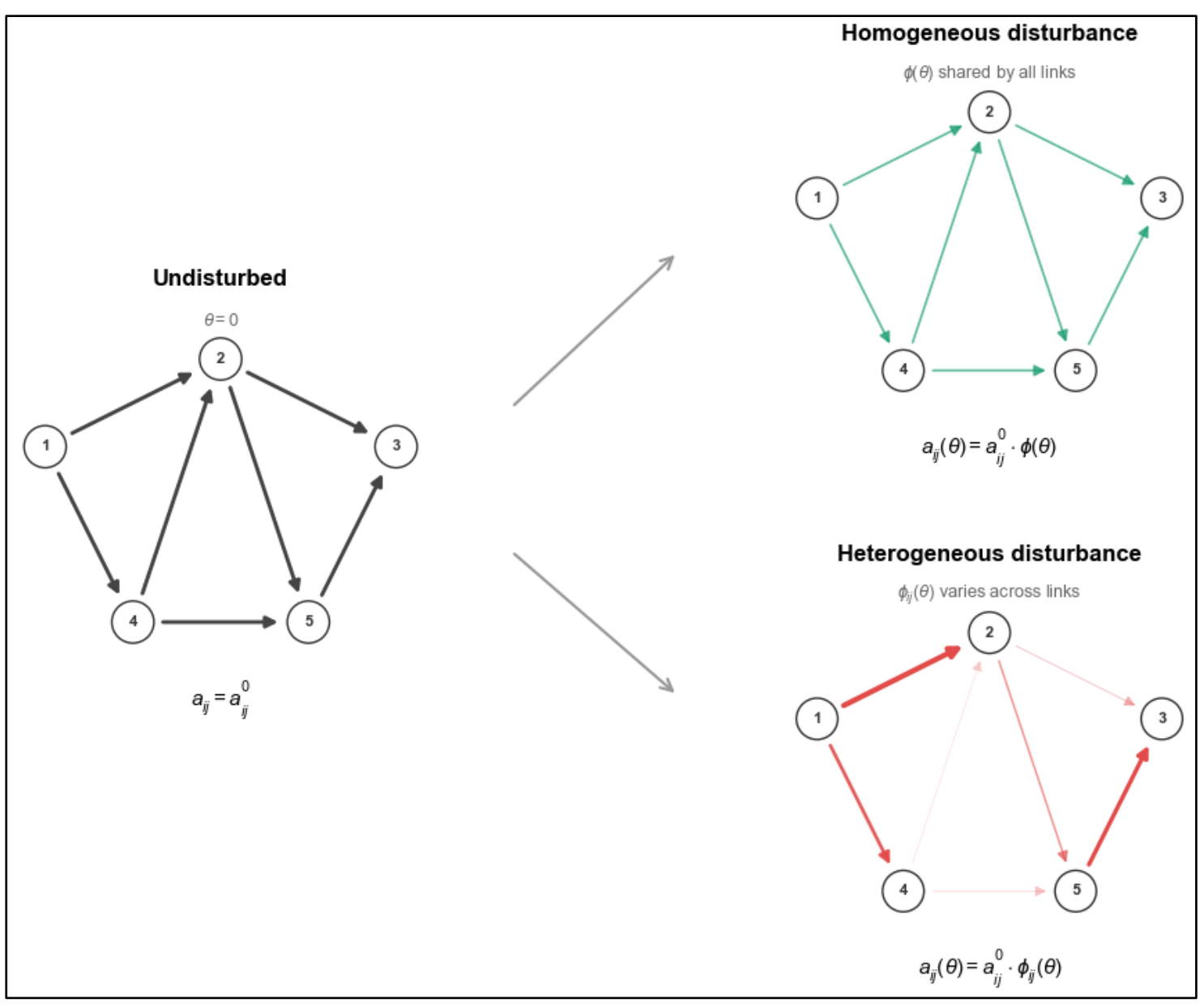

Figure 1. Transport network representation under structural disturbance.

*2.2 Propagation operator and state evolution*

To describe propagation dynamics on transport networks under structural disturbance, this study maps the transport network onto a quantum percolation system. Under this mapping, each node $i \in V$ corresponds to a basis state $|i\rangle$ and the connection strength $a_{ij}(\theta)$ serves as the propagation coupling coefficient between nodes (Mülken and Blumen, 2011).

The propagation operator is:

$$H(\theta) = \sum_{(i,j)\in E} a_{ij}(\theta)|i\rangle\langle j| \tag{2}$$

This operator embeds structural disturbance into the propagation coupling structure through the spatial distribution of $a_{ij}(\theta)$.

The propagation state is defined as a state vector $P(G(\theta),t)$ on the node basis space. It describes how the effect of a disturbance originating from a given initial configuration spread across the network over time. Its time evolution is generated by $H(\theta)$ and satisfies the following dynamical equation:

$$\mathrm{i}\frac{d}{dt}P(G(\theta),t) = H(\theta)P(G(\theta),t) \tag{3}$$

where $\mathrm{i}$ denotes the imaginary unit. When $H(\theta)$ does not vary with time, the formal solution of Eq. (3) is:

$$P(G(\theta),t) = U_\theta(t)P(G(\theta),0) \tag{4}$$

where

$$U_\theta(t) = \exp(-\mathrm{i}H(\theta)t) \tag{5}$$

is the propagation evolution operator generated by $H(\theta)$, and $P(G(\theta),0)$ is the initial propagation state.

When the propagation operator $H(\theta)$ satisfies the Hermitian condition, it admits the spectral decomposition:

$$H(\theta) = \sum_k \lambda_k(\theta)|\psi_k(\theta)\rangle\langle\psi_k(\theta)| \tag{6}$$

where $\lambda_k(\theta)$ and $|\psi_k(\theta)\rangle$ denote the $k$th eigenvalue and the corresponding eigenvector. Under this spectral representation, the time evolution of the propagation state can be expanded in modal superposition form as:

$$P(G(\theta),t)=\sum_k e^{-i\lambda_k(\theta)t}\langle\psi_k(\theta),P(G(\theta),0)\rangle|\psi_k(\theta)\rangle \tag{7}$$

The long-term propagation behaviour is jointly determined by the spectral modal structure of the propagation operator and the projection weights of the initial state onto each mode.

In directed transport networks, the connection strength matrix $A=[a_{ij}]$ is generally asymmetric, and $H(\theta)$ does not necessarily satisfy the Hermitian condition. Following the Hermitian adjacency matrix construction for directed graphs (Guo and Mohar, 2017), this study symmetrises $H(\theta)$ as:

$$\overline{H}(\theta)=\frac{1}{2}\left(H(\theta)+H(\theta)^{\dagger}\right) \tag{8}$$

where $H(\theta)^{\dagger}$ denotes the conjugate transpose of $H(\theta)$. The operator $\overline{H}(\theta)$ is Hermitian, and the spectral decomposition and evolution equations established above hold on $\overline{H}(\theta)$. For notational simplicity, $H(\theta)$ is used in subsequent sections to denote the symmetrised propagation operator.

*2.3 DPC*

To quantify the effective spreading range of the propagation state across the node space, this study defines the participation index.

$$\Pi(\theta,t)=\frac{\left(\sum_{i\in V}|P_i(G(\theta),t)|^2\right)^2}{\sum_{i\in V}|P_i(G(\theta),t)|^4} \tag{9}$$

where $P_i(G(\theta),t)$ denotes the component of the propagation state at node $i$. The value of $\Pi(\theta,t)$ represents the effective number of nodes over which the propagation state spreads. When $\Pi(\theta,t)=N$, the propagation state distributes uniformly across all nodes. When $\Pi(\theta,t)=1$, the propagation state concentrates on a single node. This index is invariant under global scaling and can distinguish between extended and localised distributions of the propagation state in node space. It has been widely used in studies of disordered systems and network propagation (De Domenico et al., 2016; Maciel et al., 2020).

To characterise propagation behaviour over long time scales, this study defines the time-averaged participation index.

$$\overline{\Pi}(\theta)=\lim_{T\to\infty}\frac{1}{T}\int_0^T\Pi(\theta,t)dt \tag{10}$$

When $\overline{\Pi}(\theta)$ grows proportionally with the network size $N$, the propagation state is extended. When $\overline{\Pi}(\theta)$ approaches a finite value independent of $N$, the propagation state is localised (Dikopoltsev et al., 2022). This study adopts $\overline{\Pi}(\theta)$ as the core measure of Dynamic Propagation Connectivity (DPC).

To analyse propagation characteristics from the perspective of spectral modes, this study defines the participation index of the $k$th propagation mode.

$$\Pi_k(\theta)=\frac{\left(\sum_{i\in V}|\psi_k^{(i)}(\theta)|^2\right)^2}{\sum_{i\in V}|\psi_k^{(i)}(\theta)|^4} \tag{11}$$

where $\psi_k^i(\theta)$ denotes the component of the eigenvector $|\psi_k(\theta)\rangle$ at node $i$. This index measures the extent to which a single propagation mode spreads across the node space and serves to distinguish extended modes from localised modes.

*2.4 Research questions*

Based on the problem formulation above, the theoretical analysis in this study addresses the following three mathematical questions.

(1) Under what conditions does $\overline{\Pi}(\theta)$ degrade, and which network structural parameters determine the degradation rate and pattern?

(2) What theoretical relationship links $\overline{\Pi}(\theta)$ with existing classical connectivity measures?

(3) On benchmark transport network topologies, what analytical characterisations describe the degradation of $\overline{\Pi}(\theta)$.

## 3. Theoretical analysis

This section establishes the theoretical properties of $\overline{\Pi}(\theta)$ derives localisation conditions and analyses its relationship with classical connectivity measures. This section develops the theoretical analysis around the mathematical questions posed in Section 2.4.

### *3.1 Fundamental properties*

This subsection establishes the basic mathematical properties of $\overline{\Pi}(\theta)$ that underpin the subsequent theoretical analysis.

**Lemma 1** (Spectral representation of $\overline{\Pi}(\theta)$ ). Let $c_k(\theta)=\langle\psi_k(\theta),P(G(\theta),0)\rangle$ . Then:

$$\overline{\Pi}(\theta)=\lim_{T\to\infty}\frac{1}{T}\int_0^T\frac{\left(\sum_{i\in V}\left|\sum_k e^{-i\lambda_k(\theta)t}c_k(\theta)\psi_k^{(i)}(\theta)\right|^2\right)^2}{\sum_{i\in V}\left|\sum_k e^{-i\lambda_k(\theta)t}c_k(\theta)\psi_k^{(i)}(\theta)\right|^4}dt \tag{12}$$

That is, $\overline{\Pi}(\theta)$ is fully determined by the spectral structure $\{\lambda_k(\theta),\psi_k^{(i)}(\theta)\}$ and the initial projection coefficients $\{c_k(\theta)\}$ . The proof is provided in Appendix A.

**Proposition 1** (Norm conservation). For all $t\geq 0$ ,

$$\|P(G(\theta),t)\|=\|P(G(\theta),0)\| \tag{13}$$

**Proof.** Since $H(\theta)$ is Hermitian, $U_\theta(t)=\exp(-iH(\theta)t)$ is unitary. Therefore $\|U_\theta(t)P(G(\theta),0)\|=\|P(G(\theta),0)\|$.

**Proposition 2** (Continuity). Let $\phi_{ij}(\theta)$ be continuous on $[0,1]$ for all $(i,j)\in E$ . Then $\overline{\Pi}(\theta)$ is continuous with respect to $\theta$ .

**Proof.** Since $\phi_{ij}(\theta)$ is continuous, the elements of $H(\theta)$ are continuous functions of $\theta$ . For Hermitian matrices, eigenvalues and eigenvectors depend continuously on matrix elements (Kato, 1966). Therefore $\lambda_k(\theta),\psi_k^{(i)}(\theta)$ and $c_k(\theta)$ are all continuous in $\theta$ , From Eq. (12), $\overline{\Pi}(\theta)$ is a composition of continuous functions and is thus continuous.

### *3.2 Localisation and degradation*

This subsection derives the conditions under which $\overline{\Pi}(\theta)$ degrades and characterises the degradation properties.

**Theorem 1** (Absence of localisation under homogeneous disturbance). Let all links share the same decay function $\phi(\theta)$ , so that $H(\theta)=\phi(\theta)H(0)$ . If the eigenvalues of $H(0)$ are distinct, then $\overline{\Pi}(\theta)=\overline{\Pi}(0)$ for all $\theta\in[0,1]$ . Homogeneous disturbance does not alter the spatial distribution of propagation modes and does not induce localisation.

**Proof.** Since $H(\theta)=\phi(\theta)H(0)$ , the eigenvectors of $H(\theta)$ coincide with those of $H(0)$ and the eigenvalues satisfy $\lambda_k(\theta)=\phi(\theta)\lambda_k(0)$ , The projection coefficients $c_k(\theta)=\langle\psi_k(\theta),P(G(\theta),0)\rangle=c_k(0)$ remain unchanged. From Eq. (12), the squared modulus $|P(G(\theta),t)|^2$ contains cross-terms of the form $e^{-i(\lambda_k(0)-\lambda_l(0))t}$ . Since the eigenvalues of $H(0)$ are distinct,

$\lambda_k(0) \neq \lambda_l(0)$ for all $k \neq l$. For any non-zero frequency $\omega$:

$$\lim_{T\to\infty} \frac{1}{T}\int_0^T e^{-\mathrm{i}\omega t} dt = 0 \tag{14}$$

Therefore, all cross-terms vanish under time averaging. The time-averaged participation index reduces to:

$$\overline{\Pi}(\theta) = \frac{\left(\sum_{i\in V}\sum_k |c_k(0)|^2 \left|\psi_k^{(i)}(0)\right|^2\right)^2}{\sum_{i\in V}\left(\sum_k |c_k(0)|^2 \left|\psi_k^{(i)}(0)\right|^2\right)^2} \tag{15}$$

This expression is independent of $\theta$, $\overline{\Pi}(\theta) = \overline{\Pi}(0)$.

**Theorem 2** (Degradation under heterogeneous disturbance). Let the decay functions $\phi_{ij}(\theta)$ vary across links. Decompose the propagation operator as $H(\theta) = \phi(\theta)H(0) + W(\theta)$, where $\phi(\theta) = \frac{1}{|E|}\sum_{(i,j)\in E}\phi_{ij}(\theta)$ is the mean decay rate and $W(\theta) = H(\theta) - \phi(\theta)H(0)$ is the heterogeneous deviation matrix. Let the eigenvalues of $H(0)$ be distinct with minimum spacing $\delta_{\min} = \min_{k\neq l}|\lambda_k(0) - \lambda_l(0)|$. By matrix perturbation theory (Kato, 1966), the $k$th eigenvector of $H(\theta)$ satisfies:

$$\psi_k(\theta) = \psi_k(0) + \sum_{l\neq k} \frac{\langle \psi_l(0)|W(\theta)|\psi_k(0)\rangle}{\phi(\theta)(\lambda_k(0) - \lambda_l(0))}\psi_l(0) + O\left(\|W(\theta)\|^2\right) \tag{16}$$

When $W(\theta) = 0$, this reduces to $\psi_k(\theta) = \psi_k(0)$ and $\overline{\Pi}(\theta) = \overline{\Pi}(0)$, recovering **Theorem 1**. When $W(\theta) \neq 0$, the correction terms redistribute the eigenvector components across the node space, and $\overline{\Pi}(\theta) < \overline{\Pi}(0)$.

The proof is provided in Appendix B.

**Proposition 3** (Monotonicity). Let the conditions of **Theorem 2** hold. For $\|W(\theta)\| < \delta_{\min}\phi(\theta)$, $\overline{\Pi}(\theta)$ decreases monotonically with respect to $\|W(\theta)\|$.

The proof is provided in Appendix C.

**Proposition 4** (Degradation rate). Let the conditions of **Theorem 2** hold, with $\eta = \|W(\theta)\|/(\delta_{\min}\phi(\theta)) < 1$. Then for each mode $k$ with non-zero coupling:

$$\Pi_k(\theta) = \Pi_k(0) - \frac{2\eta\sum_{i\in V} p_{k,i}(0) r_{k,i}}{\|\mathbf{p}_k(0)\|_2^4} - \eta^2 \cdot \frac{\sum_{i\in V} r_{k,i}^2}{\|\mathbf{p}_k(0)\|_2^4} + O(\eta^3) \tag{17}$$

where $r_{k,i}$ is defined by Eq. (C3). The $\eta^2$ term is strictly negative, confirming that the degradation of each modal participation index is at least of order $O(\eta^2)$.

The proof is provided in Appendix D.

*3.3 Comparison with classical measures*

This subsection examines the theoretical relationship between $\overline{\Pi}(\theta)$ and classical connectivity measures.

**Proposition 5** (Relationship with algebraic connectivity). Let $L(\theta)$ denote the Laplacian matrix of the network $G(\theta)$ and let $\lambda_2(\theta)$ denote its second smallest eigenvalue, known as the algebraic connectivity (Fiedler, 1973). When $\lambda_2(\theta) = 0$, the network is topologically disconnected and $\overline{\Pi}(\theta) \leq |S_{\max}|$, where $|S_{\max}|$ is the size of the largest connected component. When $\lambda_2(\theta) > 0$, the network remains topologically connected, but $\overline{\Pi}(\theta)$ can take any value in $[1, N]$. That is, $\lambda_2(\theta) > 0$ does not guarantee that $\overline{\Pi}(\theta)$ remains at a high level.

The proof is provided in Appendix E.

**Theorem 3** (Separation from classical connectivity measures). There exists a family of connected networks on which $\lambda_2(\theta)$ remains bounded away from zero while $\overline{\Pi}(\theta)$ degrades from $O(N)$ to $O(1)$. Specifically, consider a large lattice graph $G_1$ with $N$ nodes connected by a single bridge link to a small lattice graph $G_2$ with $c$ nodes, where $c$ is a fixed constant independent of $N$. Under heterogeneous disturbance, apply strong disturbance to the internal links of $G_1$ while keeping $G_2$ and the bridge link undisturbed. The bridge link maintains topological connectivity, so $\lambda_2(\theta) > 0$. As shown in Appendix F, the eigenvectors of $H(\theta)$ localise toward $G_2$, and $\overline{\Pi}(\theta)$ degrades from $O(N)$ to $O(c) = O(1)$.

The proof is provided in Appendix F.

*3.4 Characterisation on network topologies*

This subsection derives an analytical expression for $\overline{\Pi}(\theta)$ on a network topology with known spectral structure. The result serves as a theoretical benchmark for the numerical verification in Section 4. The theoretical results in Sections 3.1-3.3 hold for arbitrary network topologies. The ring graph is selected here because its eigenvalues and eigenvectors have closed-form expressions that permit exact analytical derivation. Deriving analytical benchmarks on simple topologies and verifying them through numerical experiments is a standard approach in quantum propagation theory (Maciel et al., 2020; Mülken and Blumen, 2011).

**Proposition 6** (Characterisation on ring graph). Consider a ring graph $C_N$ with odd $N$. The $N$ nodes form a cycle, and each node connects to its two neighbours with uniform initial connection strength $a_{ij}^0 = 1$. The propagation operator $H(0)$ of $C_N$ has closed-form eigenvalues and eigenvectors (Mülken and Blumen, 2011):

$$\lambda_k(0) = 2\cos\frac{2\pi k}{N}, k = 0,1,\ldots,N-1 \tag{18}$$

The corresponding eigenvectors are:

$$\psi_k^{(j)}(0) = \frac{1}{\sqrt{N}} e^{i2\pi kj/N}, j \in V \tag{19}$$

When $N$ is odd, for $k \neq 0$, the eigenvalues have pairwise degeneracy $\lambda_k(0) = \lambda_{N-k}(0)$. Since all eigenvector components satisfy $\left|\psi_k^{(j)}(0)\right|^2 = 1/N$, the modal participation index computation is unaffected by this degeneracy. The modal participation index is therefore:

$$\Pi_k(0) = \frac{1}{\sum_{i\in V}\left|\psi_k^{(i)}(0)\right|^4} = \frac{1}{N\cdot(1/N)^2} = N \tag{20}$$

This gives $\overline{\Pi}(0) = N$.

Under homogeneous disturbance, by **Theorem 1**, $\overline{\Pi}(\theta) = N$.

Under heterogeneous disturbance, each link has an independent decay function $\phi_{ij}(\theta)$. By **Proposition 4**, the degradation of each modal participation index is:

$$\Pi_k(\theta) = N - \eta^2 N^2 \sum_{i\in V} r_{k,i}^2 + O(\eta^3) \tag{21}$$

where $r_{k,i}$ is defined by Eq. (C3) and $\eta = \|W(\theta)\| / (\delta_{\min}\phi(\theta))$. The degradation scales with $N^2$, indicating that larger ring graphs experience faster degradation of modal participation indices under the same heterogeneous disturbance strength.

The proof is provided in Appendix G.

## 4. Numerical experiments

*4.1 Experimental setup*

This subsection describes the computational procedure, disturbance configuration schemes, and test networks used in the numerical experiments.

(1) Computational procedure. Given a network $G=(V,E,A)$ and disturbance intensity parameter $\theta$, the computation of $\overline{\Pi}(\theta)$ proceeds as follows. First, compute the connection strength under disturbance for each link by Eq. (1), and construct the disturbed connection strength matrix $A(\theta)$. For directed networks, apply the Hermitian symmetrisation in Eq. (8) to obtain the propagation operator $H(\theta)$. Perform eigenvalue decomposition on $H(\theta)$ to obtain eigenvalues $\lambda_k(\theta)$ and eigenvectors $|\psi_k(\theta)\rangle$. Select an initial propagation state $|\psi(0)\rangle$ and compute the projection coefficients $c_k(\theta)=\langle\psi_k(\theta)|\psi(0)\rangle$. Finally, compute $\overline{\Pi}(\theta)$ by Eq. (12). Figure 2 illustrates the complete computational procedure.

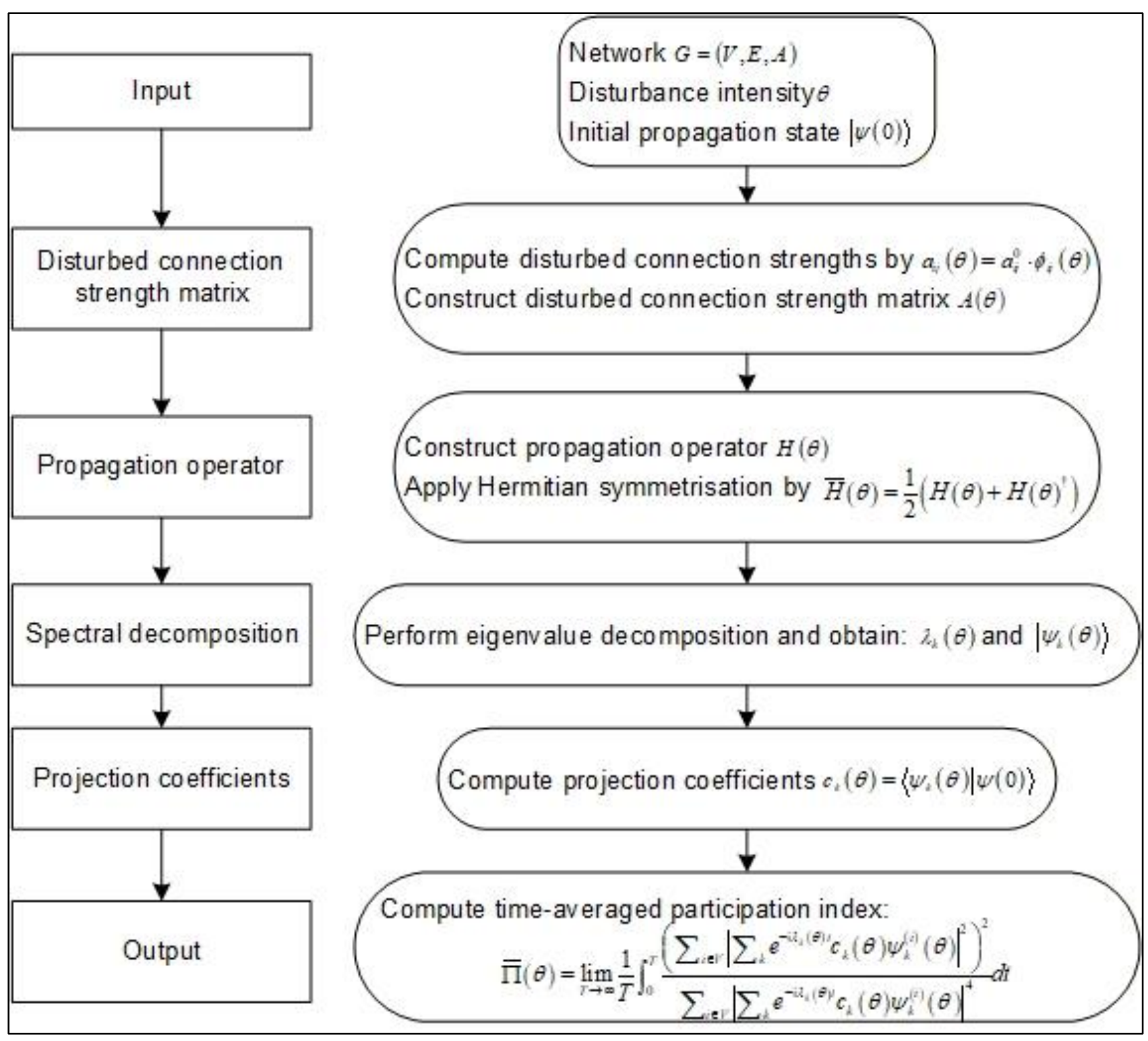

Figure 2. Computational procedure for the time-averaged participation index.

(2) Disturbance configuration schemes. The numerical experiments consider two types of disturbance. Under homogeneous disturbance, all links share the same decay function $\phi(\theta)$, set as $\phi(\theta)=1-\theta$. Under heterogeneous disturbance, each link has an independently generated decay function. Specifically, for each link $(i,j)$, set $\phi_{ij}(\theta)=1-\theta\cdot\varepsilon_{ij}$, where $\varepsilon_{ij}$ is a random variable independently sampled from the uniform distribution $U(0,1)$. This setting ensures that the decay varies across links and satisfies the four conditions on the decay function in Section 2.1. The disturbance intensity parameter takes values in $\theta\in[0,0.9]$ to avoid complete link disconnection that would cause eigenvalue degeneracy. For each heterogeneous disturbance configuration, 100 independent random samples are generated and $\overline{\Pi}(\theta)$ is averaged to eliminate random fluctuations.

(3) Initial propagation state. The initial propagation state is set as the uniform distribution, $|\psi(0)\rangle = \frac{1}{\sqrt{N}}\sum_{i\in V}|i\rangle$, so that the initial projection coefficients do not favour any particular mode. To verify the robustness of the conclusions, single-node initial states $|\psi(0)\rangle = |i\rangle$ (with randomly selected node $i$) are also tested. The qualitative conclusions remain consistent with those obtained from the uniform initial state.

The numerical experiments are conducted on the following networks.

(1) Ring graph $C_{51}$, constructed directly according to the definition in **Proposition 6**. This network serves to verify the mathematical correctness of the theoretical formulae by comparison with the analytical results of **Proposition 6.** $N = 51$ is chosen as an odd number to satisfy the condition of **Proposition 6**.

(2) Sioux Falls network, comprising 24 nodes and 76 links. This is a widely used small-scale benchmark network in transport network research.

(3) Anaheim network, comprising 416 nodes and 914 links. A medium-scale transport benchmark network.

(4) Chicago Sketch network, comprising 933 nodes and 2950 links. A large-scale transport benchmark network used to verify the applicability of the theoretical conclusions on large networks.

The above three standard benchmark transport networks are obtained from Transportation Networks for Research/Transportation Network Test Problems (TNTP) (Transportation Networks for Research Core Team, n.d.).

*4.2 Verification on ring graph*

This subsection performs numerical computations on the ring graph $C_{51}$ and compares the results with the analytical expressions in Section 3.4 to verify the mathematical correctness of the theoretical formulae.

*4.2.1 Undisturbed state*

At $\theta = 0$, the propagation operator $H(0)$ of $C_{51}$ is constructed and decomposed into eigenvalues $\lambda_k(0) = 2\cos\frac{2\pi k}{51}$ and corresponding eigenvectors. All eigenvector components have equal modulus $\left|\psi_k^{(j)}(0)\right|^2 = 1/51$, yielding modal participation indices $\Pi_k(0) = 51$ for all modes. The numerical computation gives $\overline{\Pi}(0) = 51$, in exact agreement with the analytical result $\overline{\Pi}(0) = N = 51$ from Eq. (20). This verifies **Proposition 6**.

*4.2.2 Comparison of homogeneous and heterogeneous disturbance*

Figure 3 shows the curves of $\overline{\Pi}(\theta)$ under homogeneous and heterogeneous disturbance as $\theta$ varies. Under homogeneous disturbance, $\overline{\Pi}(\theta) = 51$ holds for all $\theta \in [0, 0.9]$, verifying **Theorem 1**. Under heterogeneous disturbance, $\overline{\Pi}(\theta) < N$ for all $\theta > 0$, verifying the degradation conclusion of **Theorem 2.** $\overline{\Pi}(\theta)$ decreases monotonically as $\theta$ increases, verifying **Proposition 3.** Both curves are smooth and continuous, verifying the continuity conclusion of **Proposition 2**. The heterogeneous disturbance curve flattens after $\theta \approx 0.3$, with $\overline{\Pi}(\theta)$ stabilising at approximately 28. This occurs because the topological symmetry of the ring graph limits the degree of localisation. Even as the disturbance intensity continues to increase, the propagation state still covers more than half of the nodes.

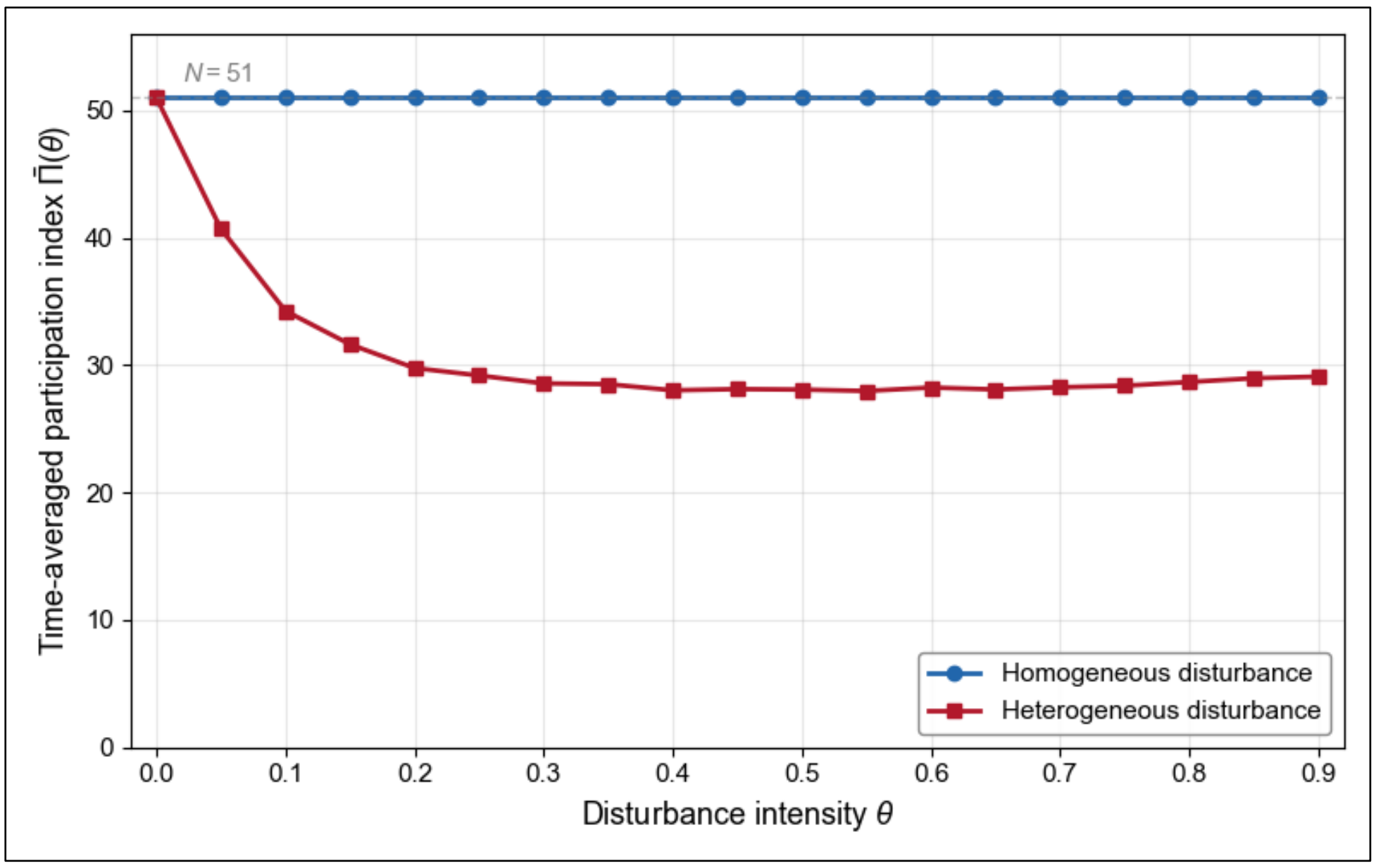

Figure 3. DPC under homogeneous and heterogeneous disturbance on ring graph.

Figure 3 provides an intuitive observation. DPC measures the number of nodes that the propagation state effectively covers. Under homogeneous disturbance, all links degrade by the same proportion. Propagation slows down but remains uniformly distributed across all nodes, so DPC stays constant at $N$ . Under heterogeneous disturbance, unequal degradation across links reshapes the propagation pattern and concentrates it towards a subset of nodes. DPC drops rapidly as a result. This indicates that disturbance variation across links drives the spatial distortion of propagation, not overall disturbance intensity.

*4.2.3 Verification of the analytical expansion*

Figure 4 compares the numerically computed DPC with the second-order analytical expansion in Eq. (21) under heterogeneous disturbance on $C_{51}$ . At $\theta = 0$ , both values equal 51. In the weak disturbance range, the two curves closely agree, with a relative error below 2%. As the disturbance intensity increases, the analytical expansion gradually overestimates DPC because the truncation at second order omits higher-order correction terms. This is consistent with the expected behaviour of perturbation expansions. Both curves share the same monotonic degradation trend, confirming that Eq. (21) correctly captures the direction and mechanism of degradation. The value of the analytical expansion lies in revealing the structural dependence of degradation on network size and disturbance heterogeneity, as established in **Proposition 4** and **Proposition 6**. Precise quantification of DPC on real transport networks relies on the numerical computation presented in Section 4.3.

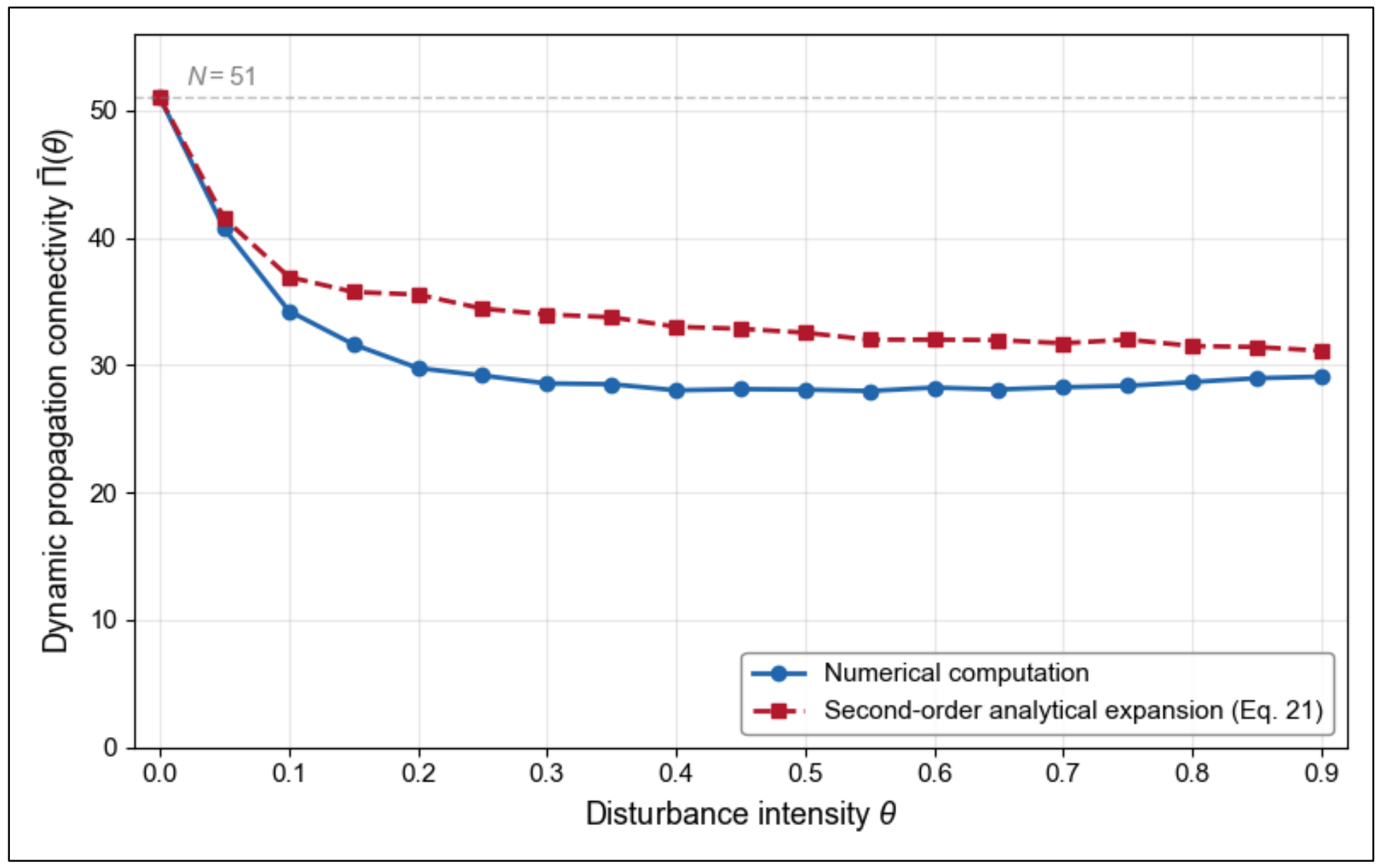

Figure 4. Comparison of numerically computed DPC with the second-order analytical expansion on ring graph.

*4.3 Numerical analysis on transport networks*

This subsection applies the computational procedure in Section 4.1 to three standard transport benchmark networks and examines whether the theoretical conclusions established in Section 3 hold on real transport network topologies. The analysis covers three aspects. They include the comparison of DPC under homogeneous and heterogeneous disturbance, the separation between DPC and algebraic connectivity, and the effect of network size on degradation.

*4.3.1 Comparison of homogeneous and heterogeneous disturbance*

Figure 5 shows the DPC $\bar{\Pi}(\theta)$ under homogeneous and heterogeneous disturbance on the Sioux Falls, Anaheim, and Chicago Sketch networks as $\theta$ varies. Under homogeneous disturbance, $\bar{\Pi}(\theta)$ remains constant on all three networks, at 16.2, 226.4, and 388.1 respectively. This verifies **Theorem 1**. Under heterogeneous disturbance, $\bar{\Pi}(\theta)$ decreases on all three networks as $\theta$ increases, verifying the degradation conclusion of **Theorem 2**. The degradation trend is monotonic overall, verifying **Proposition 3**. Both curves are continuous and smooth on all three networks, consistent with **Proposition 2**. The undisturbed values $\bar{\Pi}(0)$ are lower than $N$ on all three networks. This occurs because real transport networks have non-uniform degree distributions. The eigenvectors of the propagation operator are not uniformly distributed across nodes even in the undisturbed state. This differs from the ring graph, where uniform degree distribution yields $\bar{\Pi}(0) = N$ .

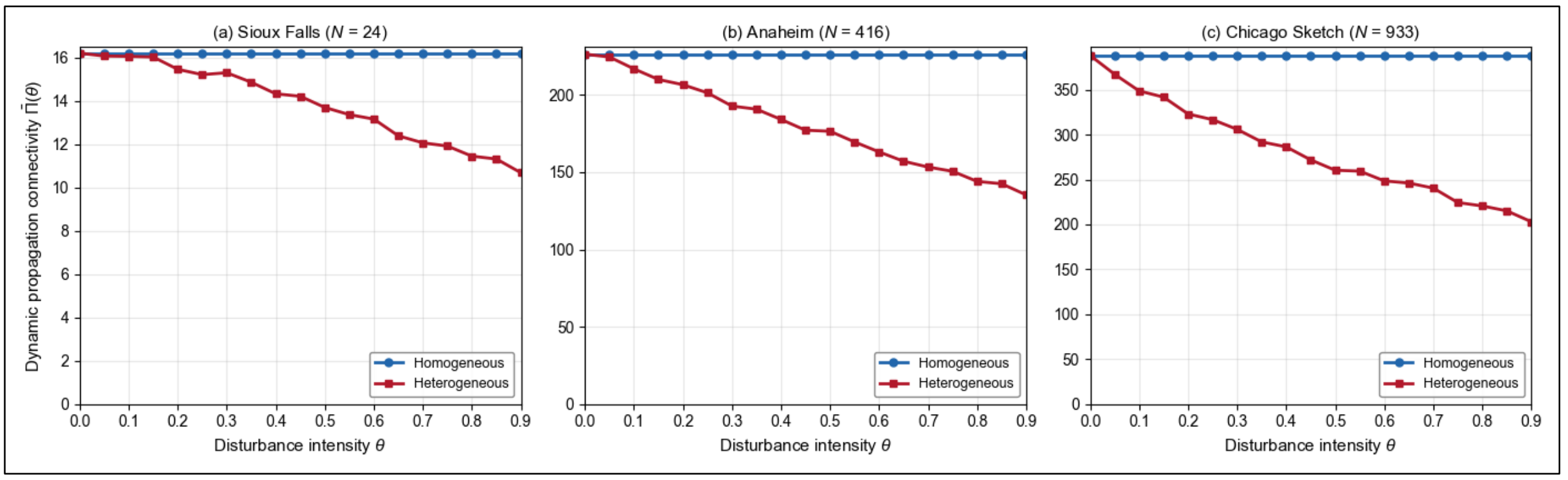

Figure 5. DPC under homogeneous and heterogeneous disturbance on transport benchmark networks.

Figure 5 confirms the same pattern observed on the ring graph in Section 4.2.2. When all links degrade by the same proportion, the propagation state slows down but its spatial distribution remains unchanged. DPC stays constant. When different links degrade by different amounts, the propagation state concentrates towards subsets of nodes where connections remain relatively strong. DPC drops as a result. The three networks differ in scale and topology, but this pattern holds consistently. The degradation is more pronounced on larger networks, where the propagation state has more room to redistribute and localise.

*4.3.2 Separation from algebraic connectivity*

Figure 6 shows the separation between DPC and algebraic connectivity under heterogeneous disturbance on the three transport benchmark networks. The left column plots both measures normalised to their undisturbed values on the same axis. On all three networks, the normalised algebraic connectivity decreases only mildly as disturbance intensity increases and remains above 0.8 throughout. This indicates that the network stays well connected in the topological sense. Over the same disturbance range, the normalised DPC drops to 0.53 to 0.67. The shaded region between the two curves shows the extent of separation. This result verifies **Proposition 5** and **Theorem 3**. Positive algebraic connectivity does not guarantee that DPC remains high. The two measures capture fundamentally different aspects of network connectivity.

The right column quantifies the separation degree, defined as the difference between the normalised algebraic connectivity and the normalised DPC. On all three networks, the separation degree rises from zero as the disturbance intensity increases. At the strongest disturbance, the separation degree reaches approximately 0.23 on Sioux Falls, 0.25 on Anaheim, and 0.30 on Chicago Sketch. Larger networks exhibit greater separation. This indicates that the gap between topological connectivity and propagation connectivity becomes more pronounced on large transport networks.

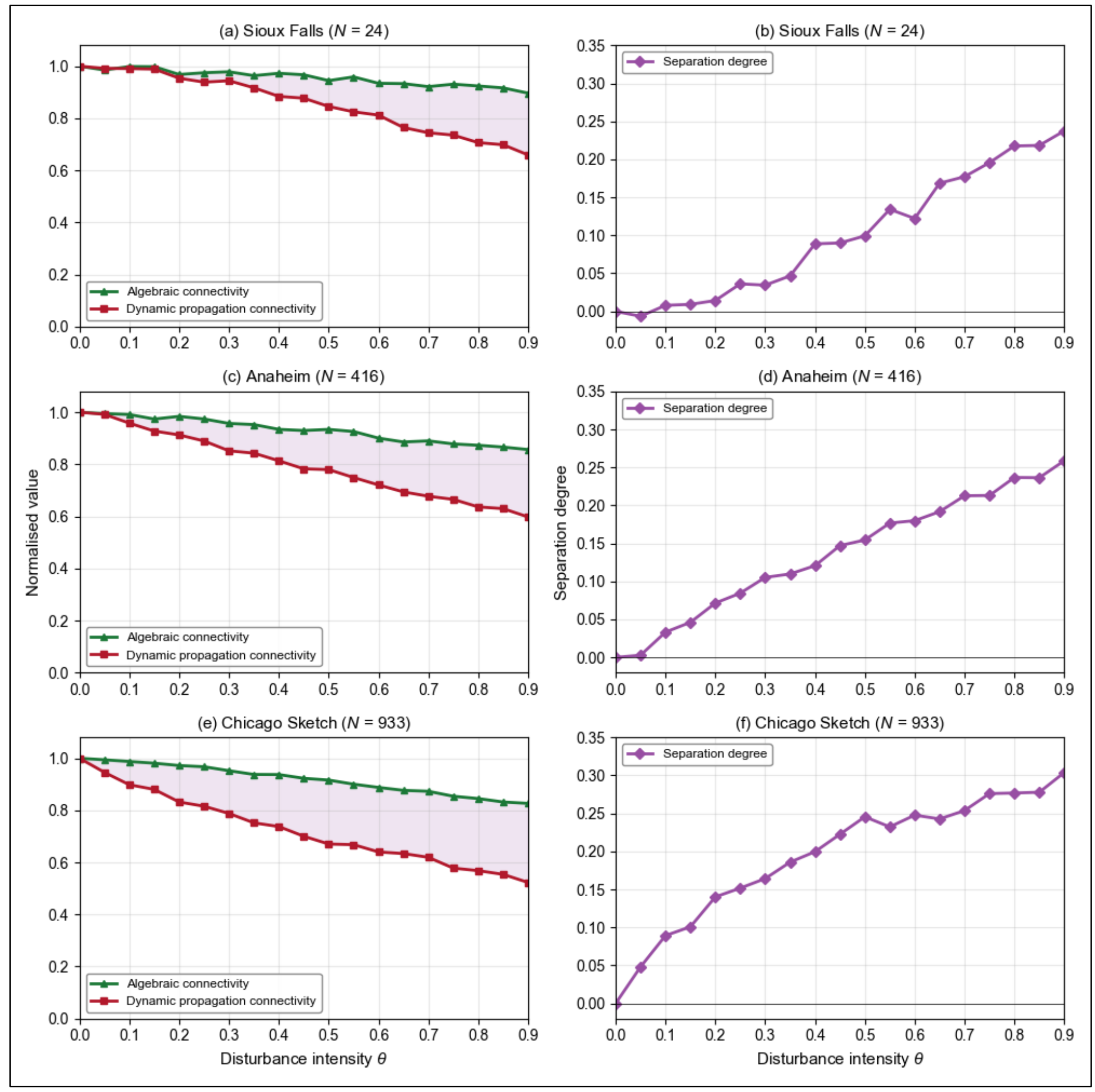

Figure 6. Separation between DPC and algebraic connectivity under heterogeneous disturbance on transport benchmark networks.

This separation has a clear physical interpretation. Algebraic connectivity measures whether every node can be reached through some path. DPC measures how effectively a disturbance signal spreads across the network in practice. A network can remain fully connected in the topological sense while its ability to propagate signals has already deteriorated. Heterogeneous disturbance creates weak links that act as bottlenecks. Paths still exist, but propagation concentrates on a subset of nodes and avoids the weakened regions. This distinction matters for transport network resilience assessment. Evaluating only topological connectivity can overestimate the functional performance of a network under disturbance. Classical percolation and topological connectivity measures detect network fragmentation but cannot capture this type of propagation degradation. The quantum percolation framework captures degradation before topological disconnection occurs.

*4.3.3 Effect of network size on degradation*

Figure 7 analyses the degradation of DPC from the perspective of network size. The left panel shows a heatmap of the normalised DPC across the three networks at varying disturbance intensities. The colour represents the ratio of DPC under disturbance to its undisturbed value, ranging from 1.0 (no degradation, green) to 0.5 (strong degradation, red). Sioux Falls shows the most gradual colour change, while Chicago Sketch shows the most rapid transition. This indicates that larger networks are more sensitive to heterogeneous disturbance.

The right panel quantifies the degradation percentage at three representative disturbance levels. Under weak disturbance, Sioux Falls degrades by approximately 5.6%, Anaheim by 14.8%, and Chicago Sketch by 21.2%. Under strong disturbance, the degradation percentages reach 34.1%, 40.2%, and 47.7% respectively. At every disturbance level, the degradation percentage increases with network size. This trend is consistent with the theoretical conclusion of **Proposition 6** on the ring graph, where degradation scales with network size. In large transport networks, the propagation state has more room to redistribute and localise, which makes them more vulnerable to heterogeneous disturbance.

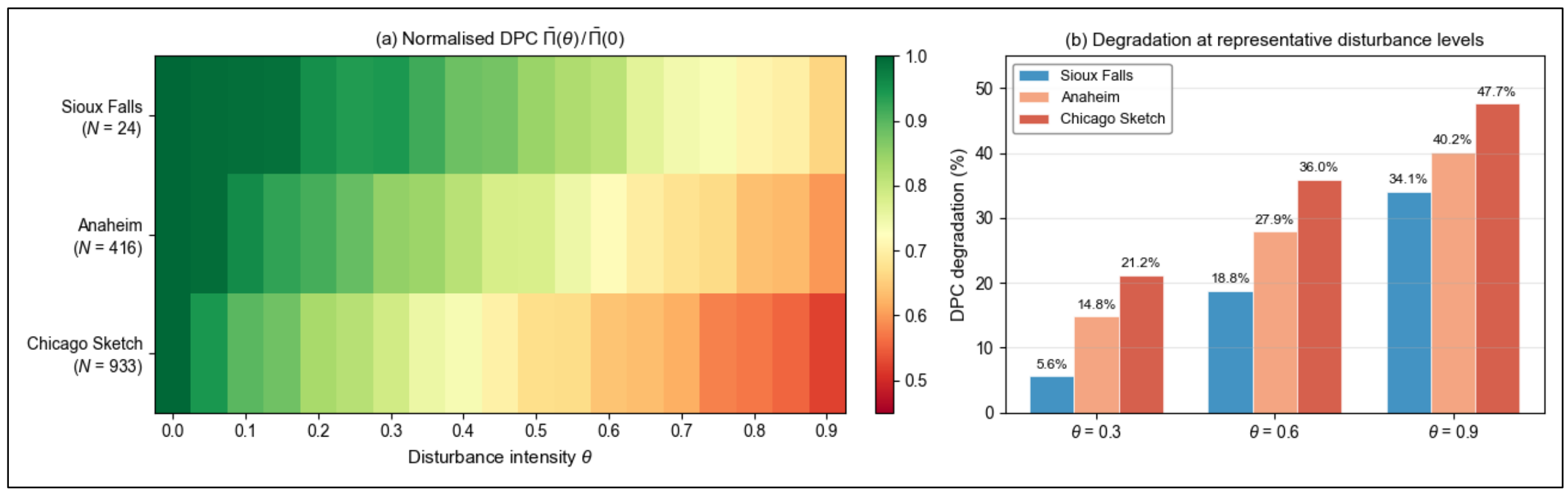

Figure 7. Effect of network size on degradation of DPC under heterogeneous disturbance.

*4.3.4 Robustness check with single-node initial state*

Figure 8 shows the distribution of DPC across all single-node initial states under heterogeneous disturbance. At each disturbance level, the propagation computation starts from every node in the network individually, and the resulting DPC values form the distribution shown by the violin plot. The box plot overlay shows the median, interquartile range, and outliers.

On all three networks, the median DPC decreases as the disturbance intensity increases, consistent with the degradation trend observed under the uniform initial state in Section 4.3.1. The distribution widens as the disturbance intensity grows. At low disturbance, different initial nodes produce similar DPC values. At high disturbance, the spread increases substantially, indicating that some nodes retain good propagation coverage while others become highly localised. This is a direct consequence of heterogeneous disturbance creating uneven propagation conditions across the network.

The dashed line marks the DPC value obtained from the uniform initial state. The median of the single-node distribution tracks closely with this reference value throughout the disturbance range. This confirms that the degradation conclusions in Sections 4.3.1 through 4.3.3 do not depend on the choice of initial propagation state. The qualitative findings hold for both uniform and single-node initial conditions.

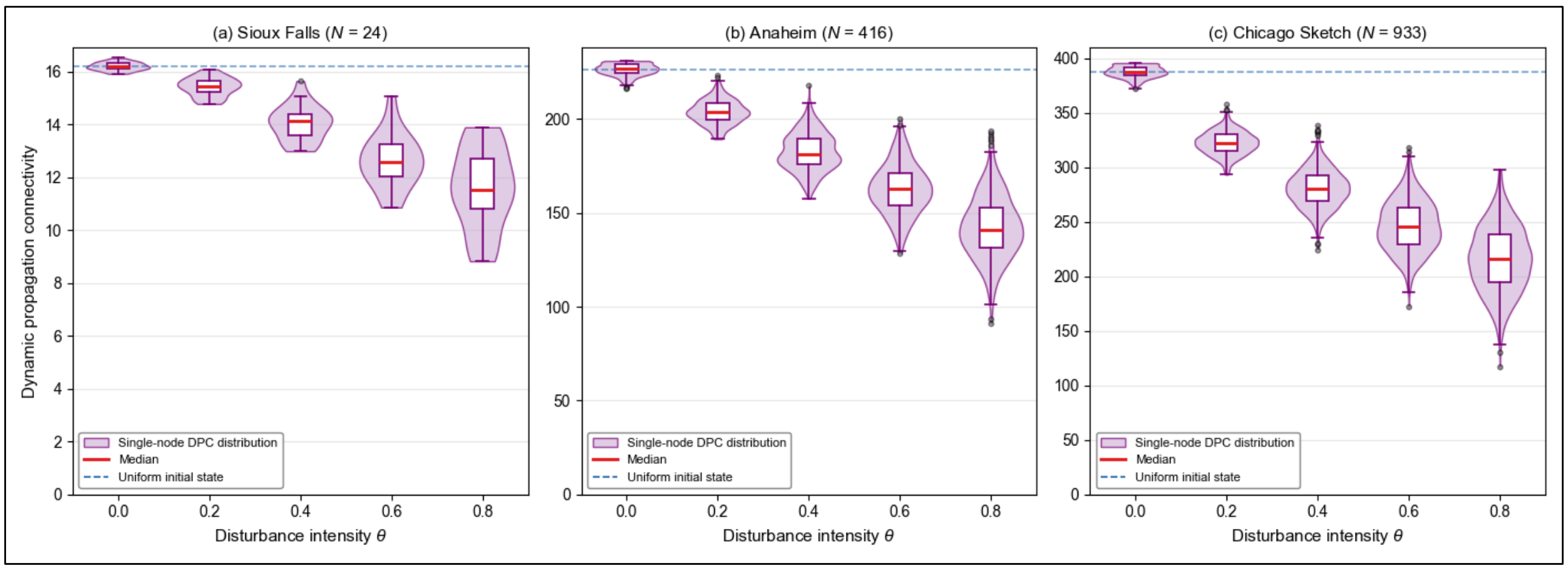

Figure 8. Distribution of DPC across single-node initial states under heterogeneous disturbance on transport benchmark networks.

*4.3.5 Modal participation index distribution*

Figure 9 presents the modal participation index distribution on the three transport benchmark networks under heterogeneous disturbance. The left column plots the modal participation index of each mode at the undisturbed state against its value under strong disturbance. The dashed diagonal line represents the case where no degradation occurs. On all three networks, most points fall below the diagonal. This confirms that heterogeneous disturbance reduces the spatial extent of individual propagation modes. The deviation from the diagonal varies across modes. Some modes retain relatively high participation indices, while others drop sharply. This indicates that heterogeneous disturbance does not affect all modes uniformly. Certain modes are more susceptible to localisation than others.

The right column shows the distribution of normalised modal participation indices at three disturbance levels. At the undisturbed state, the distribution concentrates in a narrow range, indicating that most modes have similar spatial extent. As disturbance intensity increases, the distribution spreads and shifts towards lower values. Under strong disturbance, a bimodal pattern emerges on Anaheim and Chicago Sketch. One group of modes retains relatively high participation indices, while a second group clusters at low values. This splitting reflects the coexistence of extended and localised modes within the same network. The pattern is consistent with the perturbation analysis in **Theorem 2**, where the magnitude of eigenvector correction depends on the coupling between each mode and the deviation matrix.

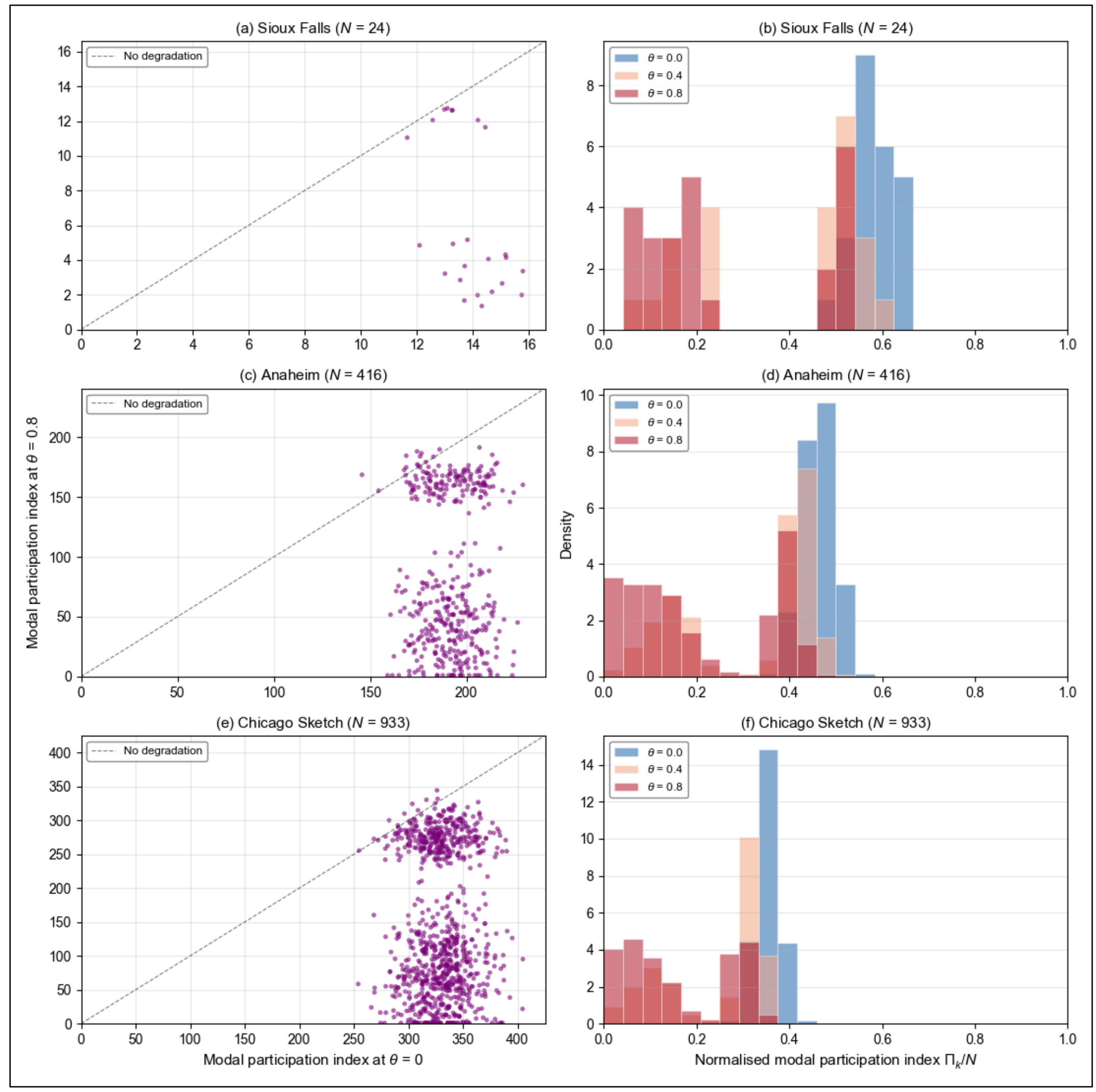

Figure 9. Modal participation index distribution under heterogeneous disturbance on transport benchmark networks.

**5. Discussion**

This paper develops a theoretical framework for analysing DPC in transport networks based on quantum percolation. The numerical experiments confirm the theoretical results across different network topologies and scales. The following discussion addresses methodological contributions, positioning within the existing literature, and limitations.

(1) Classical approaches to transport network connectivity analysis rely on two main tools. The first is topological connectivity analysis based on graph theory, which examines whether paths exist between nodes. The second is percolation theory, which tracks the evolution of connected components under random failures. Both approaches treat connectivity as a structural property determined by network topology. They do not model how signals or flows propagate through the network. The quantum percolation mapping proposed in this paper translates disturbance-induced connection decay into spectral changes of the propagation operator and defines a propagation-based connectivity measure. The separation phenomenon observed in Section 4.3.2 demonstrates that topological connectivity and propagation connectivity describe different aspects of network performance. A topologically intact network can exhibit substantial propagation degradation. This distinction carries practical implications for transport network resilience assessment.

(2) The theoretical results of this paper can be situated within three lines of existing research. In the classical percolation literature, Li et al. (2015) proposed the concept of traffic percolation and demonstrated how urban networks transition from global connectivity to fragmentation through evolving critical bottlenecks. Hamedmoghadam et al. (2021) introduced heterogeneous demand flows and capacities into the percolation framework and revealed how heterogeneity influences network connectivity vulnerability. These studies analyse connectivity degradation through the giant connected component and treat connectivity as a binary topological property. They cannot detect functional degradation before the network fragments. The separation results in Section 4.3.2 provide direct theoretical evidence for this limitation. In the quantum propagation literature, Anderson (1958) established the theory of localisation in disordered systems, showing that wave functions can become spatially confined under sufficient structural disorder. Schubert et al. (2005) extended this analysis to quantum percolation on lattice structures and characterised localisation transitions. Mülken and Blumen (2011) provided a comprehensive theoretical treatment of continuous-time quantum walks on complex networks and identified localisation as a central feature of propagation in disordered networks. This paper extends their framework to transport network topologies and provides rigorous analytical conditions under which localisation occurs (**Theorem 2**) and does not occur (**Theorem 1**). The distinction between homogeneous and heterogeneous disturbance formalises a phenomenon that previous quantum percolation studies observed numerically but did not characterise analytically. In the transport vulnerability literature, Cats and Jenelius (2018) showed that partial capacity degradation can reduce public transport network performance without causing topological disconnection. Mattsson and Jenelius (2015) reviewed existing vulnerability and resilience frameworks and highlighted the need for analytical approaches that capture functional degradation beyond path accessibility. The DPC framework proposed in this paper responds to this need by providing a spectral measure that quantifies propagation degradation continuously as disturbance intensity increases.

(3) The analysis in this paper relies on several simplifying assumptions. First, the numerical experiments set initial

connection strengths to uniform values. Real transport networks have connection strengths that vary with road capacity, lane count, and speed limits. Second, the disturbance model is static. The disturbance intensity does not change over time during the analysis. Real natural hazards often exhibit spatiotemporal evolution. Third, the Hermitian symmetrisation of the propagation operator converts directed networks into undirected networks and loses the directionality of traffic flows. These simplifications ensure the feasibility of theoretical analysis but require adjustment when applying the framework to specific scenarios.

**6. Conclusion**

This paper establishes a theoretical framework for analysing dynamic propagation connectivity in transport networks based on quantum percolation theory. The framework is supported by rigorous mathematical proofs and verified through numerical experiments. The main findings are as follows.

First, the spatial distribution of disturbance determines whether propagation connectivity degrades. When all links decay by the same proportion, the propagation pattern remains unchanged, and propagation connectivity stays constant. When different links decay by different amounts, the propagation state concentrates towards a subset of nodes and propagation connectivity decreases. The degradation rate depends on the minimum eigenvalue spacing of the undisturbed propagation operator and the strength of the heterogeneous deviation. Networks with closely spaced eigenvalues degrade faster under the same disturbance intensity. This means that the real threat to transport network function comes from the spatial unevenness of disturbance rather than disturbance intensity alone.

Second, degradation of propagation connectivity occurs before topological disconnection. The numerical experiments show that propagation connectivity has already dropped substantially while algebraic connectivity remains at a high level. Relying solely on topological connectivity to assess network resilience overestimates the functional performance of the network under disturbance. Propagation connectivity provides a more sensitive early warning indicator.

Third, network size amplifies the degradation effect of heterogeneous disturbance. Under the same disturbance conditions, larger networks exhibit greater degradation in propagation connectivity. The propagation state has more room to redistribute and localise in large networks, making large transport networks more vulnerable to heterogeneous disturbance.

Fourth, this paper derives an analytical expression for propagation connectivity on the ring graph and obtains a second-order approximation of degradation through perturbation expansion. The numerical experiments confirm the agreement between the analytical formula and numerical computation. These analytical results provide quantitative tools for understanding the degradation mechanism and lay the groundwork for deriving similar formulae on other regular topologies.

Future research can extend this work in several directions. First, the static disturbance model can be generalised to a time-varying disturbance model, enabling the framework to describe the temporal evolution of disturbance intensity during natural hazards. Second, the propagation connectivity measure can be embedded into an optimisation model to identify critical links that have the greatest impact on propagation connectivity, supporting reinforcement and resource allocation decisions for transport networks. Third, the theoretical framework can be combined with real hazard scenarios to validate the practical value of propagation connectivity on real transport network data.

**APPENDIX**

**Appendix A. Proof of Lemma 1**

By Eq. (6), the propagation operator admits the spectral decomposition:

$$H(\theta)=\sum_k \lambda_k(\theta)\left|\psi_k(\theta)\right\rangle\left\langle\psi_k(\theta)\right| \tag{A1}$$

The propagation evolution operator in Eq. (5) can then be written as:

$$U_\theta(t)=\exp\left(-\mathrm{i}H(\theta)t\right)=\sum_k e^{-\mathrm{i}\lambda_k(\theta)t}\left|\psi_k(\theta)\right\rangle\left\langle\psi_k(\theta)\right| \tag{A2}$$

Applying $U_\theta(t)$ to the initial state $P(G(\theta),0)$ and projecting onto node $i$ yields:

$$P_i(G(\theta),t)=\left\langle i\middle|U_\theta(t)\middle|P(G(\theta),0)\right\rangle=\sum_k e^{-\mathrm{i}\lambda_k(\theta)t}c_k(\theta)\psi_k^{(i)}(\theta) \tag{A3}$$

where $c_k(\theta)=\left\langle\psi_k(\theta),P(G(\theta),0)\right\rangle$ The squared modulus at node $i$ is:

$$\left|P_i(G(\theta),t)\right|^2=\left|\sum_k e^{-\mathrm{i}\lambda_k(\theta)t}c_k(\theta)\psi_k^{(i)}(\theta)\right|^2=\sum_{k,l}e^{-\mathrm{i}\left(\lambda_k(\theta)-\lambda_l(\theta)\right)t}c_k(\theta)\overline{c_l(\theta)}\psi_k^{(i)}(\theta)\overline{\psi_l^{(i)}(\theta)} \tag{A4}$$

Substituting Eq. (A4) into the numerator and denominator of the participation index defined in Eq. (9) gives $\Pi(\theta,t)$ as a function of the spectral quantities $\left\{\lambda_k(\theta),\psi_k^{(i)}(\theta),c_k(\theta)\right\}$ and the oscillatory terms $e^{-\mathrm{i}\left(\lambda_k(\theta)-\lambda_l(\theta)\right)t}$. Applying the time-averaging operation in Eq. (10) to $\Pi(\theta,t)$ yields:

$$\overline{\Pi}(\theta)=\lim_{T\to\infty}\frac{1}{T}\int_0^T\frac{\left(\sum_{i\in V}\left|\sum_k e^{-\mathrm{i}\lambda_k(\theta)t}c_k(\theta)\psi_k^{(i)}(\theta)\right|^2\right)^2}{\sum_{i\in V}\left|\sum_k e^{-\mathrm{i}\lambda_k(\theta)t}c_k(\theta)\psi_k^{(i)}(\theta)\right|^4}dt \tag{A5}$$

Since the only quantities entering this expression are $\lambda_k(\theta),\psi_k^{(i)}(\theta),c_k(\theta)$, $\overline{\Pi}(\theta)$ is fully determined by the spectral structure and the initial projection coefficients.

**Appendix B. Proof of Theorem 2**

Write $H(\theta)=H_0(\theta)+W(\theta)$, where $H_0(\theta)=\phi(\theta)H(0)$. The eigenvalues and eigenvectors of $H_0(\theta)$ are $\lambda_k^{(0)}(\theta)=\phi(\theta)\lambda_k(0)$ and $\left|\psi_k(0)\right\rangle$ respectively.

Treat $W(\theta)$ as a perturbation to $H_0(\theta)$. Since the eigenvalues of $H(0)$, the eigenvalues of $H_0(\theta)$ are also distinct for $\phi(\theta)>0$, with minimum spacing $\phi(\theta)\delta_{\min}$.

By non-degenerate perturbation theory (Kato, 1966; Bhatia, 2013), the first-order correction to the $k$th eigenvector is:

$$\left|\psi_k^{(1)}\right\rangle=\sum_{l\neq k}\frac{\left\langle\psi_l(0)\middle|W(\theta)\middle|\psi_k(0)\right\rangle}{\lambda_k^{(0)}(\theta)-\lambda_l^{(0)}(\theta)}\left|\psi_l(0)\right\rangle \tag{B1}$$

Substituting $\lambda_k^{(0)}(\theta)-\lambda_l^{(0)}(\theta)=\phi(\theta)\left(\lambda_k(0)-\lambda_l(0)\right)$ into Eq. (B1) gives:

$$\left|\psi_k^{(1)}\right\rangle=\sum_{l\neq k}\frac{\left\langle\psi_l(0)\middle|W(\theta)\middle|\psi_k(0)\right\rangle}{\phi(\theta)\left(\lambda_k(0)-\lambda_l(0)\right)}\left|\psi_l(0)\right\rangle \tag{B2}$$

The perturbed eigenvector up to first order is:

$$\left|\psi_k(\theta)\right\rangle=\left|\psi_k(0)\right\rangle+\left|\psi_k^{(1)}\right\rangle+O\left(\left\|W(\theta)\right\|^2\right) \tag{B3}$$

which yields Eq. (16). The component of the perturbed eigenvector at node $i$ is:

$$\psi_k^{(i)}(\theta)=\psi_k^{(i)}(0)+\sum_{l\neq k}\frac{\left\langle\psi_l(0)\middle|W(\theta)\middle|\psi_k(0)\right\rangle}{\phi(\theta)\left(\lambda_k(0)-\lambda_l(0)\right)}\psi_l^{(i)}(0)+O\left(\left\|W(\theta)\right\|^2\right) \tag{B4}$$

The squared modulus at node $i$ is:

$$\left|\psi_k^{(i)}(\theta)\right|^2 = \left|\psi_k^{(i)}(0)\right|^2 + 2\,\mathrm{Re}\left[\overline{\psi_k^{(i)}(0)}\sum_{l\neq k}\frac{\langle\psi_l(0)|W(\theta)|\psi_k(0)\rangle}{\phi(\theta)\left(\lambda_k(0)-\lambda_l(0)\right)}\psi_l^{(i)}(0)\right] + O\left(\|W(\theta)\|^2\right) \tag{B5}$$

Define $\mathbf{p}_k(0) = \left(\left|\psi_k^{(1)}(0)\right|^2, \ldots, \left|\psi_k^{(N)}(0)\right|^2\right)^{\mathrm{T}}$ and $\mathbf{p}_k(\theta) = \left(\left|\psi_k^{(1)}(\theta)\right|^2, \ldots, \left|\psi_k^{(N)}(\theta)\right|^2\right)^{\mathrm{T}}$ as the node weight distributions of the $k$th eigenvector before and after perturbation. By Eq. (B5), $\mathbf{p}_k(\theta) = \mathbf{p}_k(0) + \varepsilon_k(\theta)$, where $\varepsilon_k(\theta)$ is the perturbation correction with $\sum_{i\in V}\varepsilon_{k,i}(\theta) = 0$. From Eq. (11), the modal participation index can be written as:

$$\Pi_k(\theta) = \frac{\|\mathbf{p}_k(\theta)\|_1^2}{\|\mathbf{p}_k(\theta)\|_2^2} \tag{B6}$$

Since $\|\mathbf{p}_k(\theta)\|_1 = \sum_{i\in V}\left|\psi_k^{(i)}(\theta)\right|^2 = 1$ is preserved by normalisation, Eq. (B6) reduces to:

$$\Pi_k(\theta) = \frac{1}{\|\mathbf{p}_k(\theta)\|_2^2} \tag{B7}$$

Expanding $\|\mathbf{p}_k(\theta)\|_2^2$ gives:

$$\|\mathbf{p}_k(\theta)\|_2^2 = \sum_{i\in V}\left(p_{k,i}(0) + \varepsilon_{k,i}(\theta)\right)^2 = \|\mathbf{p}_k(0)\|_2^2 + 2\sum_{i\in V} p_{k,i}(0)\varepsilon_{k,i}(\theta) + \sum_{i\in V}\varepsilon_{k,i}(\theta)^2 \tag{B8}$$

The quadratic term $\sum_{i\in V}\varepsilon_{k,i}(\theta)^2 > 0$ when $W(\theta) \neq 0$. For the cross-term, by the Cauchy-Schwarz inequality:

$$\left|\sum_{i\in V} p_{k,i}(0)\varepsilon_{k,i}(\theta)\right| \leq \|\mathbf{p}_k(0)\|_2 \|\varepsilon_k(\theta)\|_2 \tag{B9}$$

When $\|W(\theta)\|$ is small relative to $\phi(\theta)\delta_{\min}$, the first-order correction satisfies $\|\varepsilon_k(\theta)\|_2 = O\left(\|W(\theta)\|^2 / \left(\delta_{\min}\phi(\theta)\right)^2\right)$, while the quadratic term satisfies $\sum_{i\in V}\varepsilon_{k,i}(\theta)^2 = O\left(\|W(\theta)\|^2 / \left(\delta_{\min}\phi(\theta)\right)^2\right)$. Since the cross-term is $O\left(\|W(\theta)\|^2 / \left(\delta_{\min}\phi(\theta)\right)^2\right)$ and the quadratic term is of the same order but strictly positive, the sum satisfies:

$$\|\mathbf{p}_k(\theta)\|_2^2 > \|\mathbf{p}_k(0)\|_2^2 \tag{B10}$$

From Eq. (B7):

$$\Pi_k(\theta) = \frac{1}{\|\mathbf{p}_k(\theta)\|_2^2} < \frac{1}{\|\mathbf{p}_k(0)\|_2^2} \tag{B11}$$

Since this holds for each mode $k$ with non-zero coupling $\langle\psi_l(0)|W(\theta)|\psi_k(0)\rangle$, and $\overline{\Pi}(\theta)$ is determined by the modal participation indices through Eq. (12), it follows that $\overline{\Pi}(\theta) < \overline{\Pi}(0)$.

When $W(\theta) = 0$, all correction terms vanish, $\psi_k(\theta) = \psi_k(0)$ and $\overline{\Pi}(\theta) = \overline{\Pi}(0)$, recovering **Theorem 1**.

**Appendix C. Proof of Proposition 3**

From Eq. (B4), the perturbation correction at node $i$ for the $k$th eigenvector is:

$$\varepsilon_{k,i}(\theta) = 2\,\mathrm{Re}\left[\overline{\psi_k^{(i)}(0)}\sum_{l\neq k}\frac{\langle\psi_l(0)|W(\theta)|\psi_k(0)\rangle}{\phi(\theta)\left(\lambda_k(0)-\lambda_l(0)\right)}\psi_l^{(i)}(0)\right] + O\left(\|W(\theta)\|^2\right) \tag{C1}$$

Define the perturbation strength parameter $\eta = \|W(\theta)\| / \left(\delta_{\min}\phi(\theta)\right)$. Under the condition $\eta < 1$ write $W(\theta) = \eta\delta_{\min}\phi(\theta)W$, where $W$ is the normalised deviation matrix with $\|W\| = 1$. Substituting into Eq. (C1), the correction term scales as:

$$\varepsilon_{k,i}(\theta) = \eta \cdot r_{k,i} + O\left(\eta^2\right) \tag{C2}$$

where

$$r_{k,i} = 2\,\mathrm{Re}\left[\overline{\psi_k^{(i)}(0)}\sum_{l\neq k}\frac{\delta_{\min}}{\lambda_k(0)-\lambda_l(0)}\langle\psi_l(0)|W|\psi_k(0)\rangle\psi_l^{(i)}(0)\right] \tag{C3}$$

is independent of $\eta$. From Eq. (B8), the squared $\ell^2$ norm of the weight distribution is:

$$\|\mathbf{p}_k(\theta)\|_2^2 = \|\mathbf{p}_k(0)\|_2^2 + 2\eta\sum_{i\in V}p_{k,i}(0)r_{k,i} + \eta^2\sum_{i\in V}r_{k,i}^2 + O(\eta^3) \tag{C4}$$

Compute the derivative with respect to $\eta$:

$$\frac{d}{d\eta}\|\mathbf{p}_k(\theta)\|_2^2 = 2\sum_{i\in V}p_{k,i}(0)r_{k,i} + 2\eta\sum_{i\in V}r_{k,i}^2 + O(\eta^2) \tag{C5}$$

The second term $2\eta\sum_{i\in V}r_{k,i}^2 > 0$ for all $\eta > 0$. For the first term, since $\sum_{i\in V}r_{k,i} = 0$, by the Cauchy-Schwarz inequality:

$$\left|\sum_{i\in V}p_{k,i}(0)r_{k,i}\right| \le \|\mathrm{p}_k(0)\|_2\|\mathbf{r}_k\|_2 \tag{C6}$$

The magnitude of the cross-term is bounded by $O(\eta^0)$, while the quadratic contribution grows as $O(\eta)$. Therefore, there exists $\eta_0 > 0$ such that for all $\eta \in (0,\eta_0)$ the quadratic term dominates and

$$\frac{d}{d\eta}\|\mathbf{p}_k(\theta)\|_2^2 > 0 \tag{C7}$$

Since $\|\mathbf{p}_k(\theta)\|_2^2$ is strictly increasing in $\eta$, from Eq. (B7):

$$\frac{d}{d\eta}\Pi_k(\theta) = -\frac{1}{\|\mathbf{p}_k(\theta)\|_2^4}\cdot\frac{d}{d\eta}\|\mathbf{p}_k(\theta)\|_2^2 < 0 \tag{C8}$$

This holds for each mode $k$ with non-zero coupling. Since $\overline{\Pi}(\theta)$ is determined by the modal participation indices through Eq. (12), $\overline{\Pi}(\theta)$ is strictly decreasing in $\eta$ for $\eta \in (0,\eta_0)$. That is, $\overline{\Pi}(\theta)$ decreases monotonically with respect to $\|W(\theta)\|$ when $\|W(\theta)\| < \delta_{\min}\phi(\theta)$.

**Appendix D. Proof of Proposition 4**

From Eq. (C4), the squared $\ell^2$ norm of the weight distribution for the $k$th mode is:

$$\|\mathbf{p}_k(\theta)\|_2^2 = \|\mathbf{p}_k(0)\|_2^2 + 2\eta\sum_{i\in V}p_{k,i}(0)r_{k,i} + \eta^2\sum_{i\in V}r_{k,i}^2 + O(\eta^3) \tag{D1}$$

From Eq. (B7), $\Pi_k(\theta) = 1/\|\mathbf{p}_k(\theta)\|_2^2$. Define $f(\eta) = \|\mathbf{p}_k(\theta)\|_2^2$ and write:

$$\Pi_k(\theta) = \frac{1}{f(\eta)} \tag{D2}$$

Expand $1/f(\eta)$ around $\eta = 0$. Let $\Delta f = f(\eta) - f(0)$. The Taylor expansion of $1/f(\eta)$ gives:

$$\frac{1}{f(\eta)} = \frac{1}{f(0)+\Delta f} = \frac{1}{f(0)} - \frac{\Delta f}{f(0)^2} + O(\Delta f^2) \tag{D3}$$

From Eq. (D1):

$$\Delta f = 2\eta\sum_{i\in V}p_{k,i}(0)r_{k,i} + \eta^2\sum_{i\in V}r_{k,i}^2 + O(\eta^3) \tag{D4}$$

Substituting Eq. (D4) into Eq. (D3) and noting that $f(0) = \|\mathbf{p}_k(0)\|_2^2$ and $1/f(0) = \Pi_k(0)$.

$$\Pi_k(\theta) = \Pi_k(0) - \frac{2\eta\sum_{i\in V}p_{k,i}(0)r_{k,i} + \eta^2\sum_{i\in V}r_{k,i}^2}{\|\mathbf{p}_k(0)\|_2^4} + O(\eta^3) \tag{D5}$$

Separating by order of $\eta$.

$$\Pi_k(\theta) = \Pi_k(0) - \frac{2\eta\sum_{i\in V}p_{k,i}(0)r_{k,i}}{\|\mathbf{p}_k(0)\|_2^4} - \eta^2\cdot\frac{\sum_{i\in V}r_{k,i}^2}{\|\mathbf{p}_k(0)\|_2^4} + O(\eta^3) \tag{D6}$$

The $\eta^2$ coefficient $\sum_{i\in V} r_{k,i}^2 > 0$ when $W(\theta)\neq 0$, since the perturbation correction $r_{k,i}$ defined by Eq. (C3) cannot vanish at all nodes simultaneously when non-zero coupling $\langle\psi_l(0)|W(\theta)|\psi_k(0)\rangle$ exists. Therefore the $\eta^2$ term provides a strictly negative contribution to $\Pi_k(\theta)$, confirming that the degradation of each modal participation index is at least of order $O(\eta^2)$.

**Appendix E. Proof of Proposition 5**

The proof proceeds in two parts.

**Part 1.** When $\lambda_2(\theta)=0$, the network $G(\theta)$ is disconnected and consists of at least two connected components $S_1, S_2, \ldots, S_m$ with $m\geq 2$. Let $S_{\max}$ denote the largest component with $|S_{\max}|<N$. The propagation operator $H(\theta)$ is block diagonal with respect to these components. Therefore, the eigenvectors of $H(\theta)$ are supported on individual components, meaning $\psi_k^{(i)}(\theta)=0$ for all nodes $i$ outside the component containing the support of $|\psi_k(\theta)\rangle$.

For any eigenvector $|\psi_k(\theta)\rangle$ supported on component $S_j$ with $|S_j|<N$, the modal participation index satisfies:

$$\Pi_k(\theta)=\frac{\left(\sum_{i\in V}\left|\psi_k^{(i)}(\theta)\right|^2\right)^2}{\sum_{i\in V}\left|\psi_k^{(i)}(\theta)\right|^4}\leq |S_j|\leq |S_{\max}| \tag{E1}$$

Since every mode is confined to a component smaller than the full network, $\Pi_k(\theta)\leq |S_{\max}|$ for all $k$ , From Eq. (12), $\overline{\Pi}(\theta)$ is determined by the modal participation indices, so $\overline{\Pi}(\theta)\leq |S_{\max}|$.

**Part 2.** When $\lambda_2(\theta)>0$, the network is topologically connected. It suffices to show that $\overline{\Pi}(\theta)$ can take any value in $[1,N]$ despite $\lambda_2(\theta)>0$.

Consider a connected network with $N$ nodes and uniform initial connection strength $a_{ij}^0=1$. Under homogeneous disturbance, by **Theorem 1**, $\overline{\Pi}(\theta)=\overline{\Pi}(0)$. For a complete graph $K_N$, the eigenvectors of $H(0)$ distribute uniformly across all nodes, giving $\Pi_k(0)=N$ for the uniform mode and $\overline{\Pi}(0)$ close to $N$. The algebraic connectivity $\lambda_2(0)=N>0$. This establishes that $\overline{\Pi}(\theta)$ can reach values close to $N$ while $\lambda_2(\theta)>0$.

Now introduce heterogeneous disturbance. By **Theorem 2**, $\overline{\Pi}(\theta)<\overline{\Pi}(0)$, and by **Proposition 3**, $\overline{\Pi}(\theta)$ decreases monotonically with increasing $\|W(\theta)\|$. Throughout this process, as long as no link is completely removed, the network remains connected and $\lambda_2(\theta)>0$.

Since $\overline{\Pi}(\theta)$ is continuous with respect to $\theta$ (**Proposition 2**), and $\|W(\theta)\|$ is a continuous function of $\theta$, by the intermediate value theorem, $\overline{\Pi}(\theta)$ takes every value in $\left[1,\overline{\Pi}(0)\right]$ as $\|W(\theta)\|$ increases. This holds while $\lambda_2(\theta)>0$ remains strictly positive. In particular, when $\|W(\theta)\|$ is large relative to $\delta_{\min}\phi(\theta)$, the eigenvectors of $H(\theta)$ become highly localised, yielding $\overline{\Pi}(\theta)$ close to 1, while the network remains topologically connected with $\lambda_2(\theta)>0$.

Therefore $\lambda_2(\theta)>0$ does not constrain the value of $\overline{\Pi}(\theta)$, and the two measures capture fundamentally different aspects of network connectivity.

**Appendix F. Proof of Theorem 3**

Construct the following family of networks. Let $G_1$ be an $n\times n$ lattice graph with $N=n^2$ nodes. Let $G_2$ be a small lattice graph with $c$ nodes, where $c$ is a fixed constant independent of $N$. Connect a node $u$ in $G_1$ to a node $v$ in $G_2$ by a single bridge link $(u,v)$, forming a connected network $G$ with $N+c$ nodes in total.

**Part 1: Lower bound on $\lambda_2(\theta)$.**

Let the internal links of $G_2$ and the bridge link $(u,v)$ remain undisturbed, with decay functions $\phi_{ij}(\theta)=1$. Since the bridge link is always present, $G$ remains topologically connected under any disturbance configuration $\theta$.

The Laplacian matrix $L(\theta)$ of $G$ can be decomposed as:

$$L(\theta)=L_1(\theta)\oplus L_2+L_b \tag{F1}$$

where $L_1(\theta)$ is the Laplacian of $G_1$ under disturbance, $L_2$ is the Laplacian of $G_2$ (undisturbed), and $L_b$ is the contribution of the bridge link. The bridge link $(u,v)$ has connection strength $a_{uv}^0>0$, and $L_b$ contributes $a_{uv}^0$ to the diagonal entries at $u$ and $v$ and $-a_{uv}^0$ to the off-diagonal entries at $(u,v)$ and $(v,u)$.

For any vector $\mathbf{x}\perp 1$, the Rayleigh quotient satisfies:

$$\frac{\mathbf{x}^{\mathrm{T}}L(\theta)\mathbf{x}}{\mathbf{x}^{\mathrm{T}}\mathbf{x}}\geq\frac{\mathbf{x}^{\mathrm{T}}L_b\mathbf{x}}{\mathbf{x}^{\mathrm{T}}\mathbf{x}} \tag{F2}$$

since $L_1(\theta)$ and $L_2$ are both positive semidefinite. The quadratic form of the bridge link is:

$$\mathbf{x}^{\mathrm{T}}L_b\mathbf{x}=a_{uv}^0(x_u-x_v)^2 \tag{F3}$$

By Fiedler's theorem (Fiedler, 1973), $\lambda_2(\theta)>0$ if and only if the graph is connected. Since the bridge link remains undisturbed and $G_2$ is undisturbed, the network $G$ remains connected regardless of the disturbance applied to the internal links of $G_1$ approach zero. The bridge link ensures that every node in $G_1$ (through $u$ ) remains connected to $G_2$.

Therefore, $\lambda_2(\theta)>0$ for all disturbance configurations $\theta$.

**Part 2: Degradation of $\overline{\Pi}(\theta)$ to $O(1)$.**

Apply strong heterogeneous disturbance to the internal links of $G_1$. Specifically, let the decay function $\phi_{ij}(\theta)$ for each link $(i,j)$ in $G_1$ vary independently and approach zero as $\theta$ increases.

Decompose the propagation operator as:

$$H(\theta)=H_0+V(\theta) \tag{F4}$$

where $H_0$ contains the internal links of $G_2$ and the bridge link (all undisturbed), and $V(\theta)$ contains the internal links of $G_1$ (under disturbance).

As the internal connection strengths of $G_1$ approach zero, $\|V(\theta)\|\to 0$ and $H(\theta)\to H_0$. The operator $H_0$ has non-zero coupling only between the internal nodes of $G_2$ and between the two endpoints of the bridge link.

The spectral structure of $H_0$ is as follows. The $N-1$ nodes of $G_1$ other than $u$ have no coupling to any other node under $H_0$. Each of these nodes supports a fully localised eigenvector with modal participation index $\Pi_k=1$. The remaining $c+1$ eigenvectors are supported on the $c$ nodes of $G_2$ together with node $u$, yielding modal participation indices $\Pi_k\leq c+1$.

Therefore, in the limit $H(\theta)\to H_0$, all modal participation indices satisfy $\Pi_k\leq c+1$. From Eq. (12), $\overline{\Pi}(\theta)$ is determined by the modal participation indices, so:

$$\overline{\Pi}(\theta) \le c+1 = O(1) \tag{F5}$$

**Part 3:** $\overline{\Pi}(0) = O(N)$**.**

In the undisturbed state, $G_1$ is an $n \times n$ lattice graph with all internal links intact. The eigenvectors of a lattice graph are plane-wave modes that distribute uniformly across all $N$ nodes, with modal participation indices $\Pi_k(0) = O(N)$. Since $G_1$ contributes the vast majority of nodes $(N \gg c)$, $\overline{\Pi}(0) = O(N)$.

Combining Parts 1-3, $\lambda_2(\theta) > 0$ holds for all $\theta$, while $\overline{\Pi}(\theta)$ degrades from $O(N)$ to $O(1)$.

**Appendix G. Proof of Proposition 6**

First derive the spectral structure of the ring graph $C_N$. The propagation operator $H(0)$ of $C_N$ is the adjacency matrix. Each node $j$ connects to nodes $j-1$ and $j+1$ (modulo $N$) with connection strength 1. $H(0)$ is a circulant matrix.

The eigenvectors of a circulant matrix are the discrete Fourier basis. Set $\psi_k^{(j)}(0) = \frac{1}{\sqrt{N}} e^{\mathrm{i}2\pi kj/N}$ and substitute into the eigenvalue equation $H(0)|\psi_k(0)\rangle = \lambda_k(0)|\psi_k(0)\rangle$. The equation at node $j$ is:

$$\psi_k^{(j-1)}(0) + \psi_k^{(j+1)}(0) = \lambda_k(0)\psi_k^{(j)}(0) \tag{G1}$$

The left-hand side expands as:

$$\frac{1}{\sqrt{N}} e^{\mathrm{i}2\pi k(j-1)/N} + \frac{1}{\sqrt{N}} e^{\mathrm{i}2\pi k(j+1)/N} = \frac{1}{\sqrt{N}} e^{\mathrm{i}2\pi kj/N}\left(e^{-\mathrm{i}2\pi kj/N} + e^{\mathrm{i}2\pi kj/N}\right) \tag{G2}$$

By Euler's formula, $e^{-\mathrm{i}2\pi kj/N} + e^{\mathrm{i}2\pi kj/N} = 2\cos\frac{2\pi k}{N}, k = 0,1,\ldots,N-1$. Therefore:

$$\lambda_k(0) = 2\cos\frac{2\pi k}{N}, k = 0,1,\ldots,N-1 \tag{G3}$$

This gives Eq. (18). The corresponding eigenvectors $\psi_k^{(j)}(0) = \frac{1}{\sqrt{N}} e^{\mathrm{i}2\pi kj/N}$ give Eq. (19).

When $N$ is odd, for $k \ne 0, \lambda_k(0) = \lambda_{N-k}(0)$, and the eigenvalues have pairwise degeneracy. The corresponding eigenvectors $|\psi_k(0)\rangle$ and $|\psi_{N-k}(0)\rangle$ are linearly independent and both satisfy $\left|\psi_k^{(j)}(0)\right|^2 = 1/N$. The modal participation index computation is therefore unaffected by this degeneracy.

Each eigenvector satisfies $\left|\psi_k^{(j)}(0)\right|^2 = 1/N$. The modal participation index is:

$$\Pi_k(0) = \frac{1}{\sum_{i\in V}\left|\psi_k^{(i)}(0)\right|^4} = \frac{1}{N\cdot(1/N)^2} = N \tag{G4}$$

From Eq. (12), $\overline{\Pi}(0) = N$. This gives Eq. (20).

Under homogeneous disturbance, by **Theorem 1**, the eigenvectors remain unchanged and $\overline{\Pi}(\theta) = \overline{\Pi}(0) = N$.

Under heterogeneous disturbance, by **Proposition 4**, for each mode $k$ with non-zero coupling:

$$\Pi_k(\theta) = \Pi_k(0) - \eta^2 \cdot \frac{\sum_{i\in V} r_{k,i}^2}{\|\mathbf{p}_k(0)\|_2^4} + O(\eta^3) \tag{G5}$$

Since $\left|\psi_k^{(i)}(0)\right|^2 = 1/N$,

$$\|\mathbf{p}_k(0)\|_2^2 = \sum_{i\in V}\left|\psi_k^{(i)}(0)\right|^4 = N\cdot\frac{1}{N^2} = \frac{1}{N} \tag{G6}$$

Therefore $\left\|\mathbf{p}_k(0)\right\|_2^4 = \frac{1}{N^2}$ , Substituting into Eq. (G5):

$$\Pi_k(\theta) = N - \eta^2 N^2 \sum_{i \in V} r_{k,i}^2 + O\left(\eta^3\right) \tag{G7}$$

This gives Eq. (21).

**Declaration of generative AI and AI-assisted technologies in the writing process**

During the preparation of this work, we employed ChatGPT solely to assist with language editing and polishing. No content was generated by AI. After using this tool, we thoroughly reviewed and edited the content as necessary and accept full responsibility for the content of the published article.